\documentclass[letterpaper,journal]{IEEEtran}
\usepackage[numbers,sort&compress]{natbib}

\usepackage{graphicx}
\usepackage{bm}
\usepackage[hidelinks]{hyperref}
\usepackage{amssymb}
\usepackage{amsthm}
\usepackage{mathtools}
\usepackage{esint}
\newcommand{\Sf}{\mathsf}
\newcommand{\E}{\mathbb{E}}
\newcommand{\R}{\mathbb{R}}
\newcommand{\bA}{\bm{A}}
\newcommand{\bB}{\bm{B}}
\newcommand{\bG}{\bm{G}}
\newcommand{\bL}{\bm{L}}
\newcommand{\bR}{\bm{R}}
\newcommand{\bS}{\bm{S}}
\newcommand{\bY}{\bm{Y}}
\newcommand{\sC}{\Sf{c}}
\newcommand{\dd}{\Sf{d}}
\newcommand{\sF}{\Sf{F}}
\newcommand{\sL}{\Sf{L}}
\newcommand{\sS}{\Sf{s}}
\newcommand{\Normal}{\mathcal{N}}
\newcommand{\MMSE}{\Sf{m}}
\newtheorem{conjecture}{Conjecture}
\newtheorem{lemma}{Lemma}
\newtheorem{theorem}{Theorem}
\newtheorem{remark}{Remark}
\begin{document}

\title{Extensive-Rank Matrix Denoising Under i.i.d. Non-Gaussian Noise}

\author{
Zhuohang He, Qun Chen, Junjie Ma, Haochuan Zhang, and Xiaojun Yuan
\thanks{Zhuohang He and Xiaojun Yuan are with the National Key Laboratory of Wireless Communications, University of Electronic Science and Technology of China, Chengdu 610056, China (e-mail: 201911220520@std.uestc.edu.cn; xjyuan@uestc.edu.cn).}%
\thanks{Qun Chen and Haochuan Zhang are with the School of Automation, Guangdong University of Technology, Guangzhou 510006, China (e-mail: c.prey.q@gmail.com; haochuan.zhang@gdut.edu.cn).}%
\thanks{Junjie Ma is with the Institute of Computational Mathematics and Scientific/Engineering Computing, Academy of Mathematics and Systems Science, Chinese Academy of Sciences, Beijing 100864, China (e-mail: majunjie@lsec.cc.ac.cn).}%
\thanks{Corresponding authors: Qun Chen, Junjie Ma, and Xiaojun Yuan.}
}

\maketitle

\begin{abstract}

We study extensive-rank matrix denoising for rotationally invariant signals corrupted by i.i.d. non-Gaussian noise. To exploit both the signal spectrum and the entrywise noise prior, we consider an orthogonal approximate message passing (OAMP) algorithm that alternates between a rotationally invariant spectral estimator and an entrywise denoiser. At each OAMP iteration, we conjecture a high-dimensional Gaussian equivalence property: the entrywise module admits an effective additive scalar Gaussian channel, whereas the spectral module behaves as if corrupted by i.i.d. Gaussian noise of matching variance. Under this conjecture, we formulate a state evolution (SE) recursion for the effective noise variances of both modules. Numerical simulations confirm that the SE predictions closely track the empirical mean-square errors. For robust principal component analysis (PCA), we apply this OAMP algorithm and characterize its recovery phase diagram via SE. Numerical results show that OAMP achieves a larger recovery region than several existing methods.

\end{abstract}


\IEEEpeerreviewmaketitle

\section{Introduction}

\IEEEPARstart{T}{he} goal of matrix denoising is to recover a structured low-rank matrix from noise-corrupted observations. Classical spiked matrix models typically focus on rank-one or fixed-rank signal components, a regime with foundational results on spectral phase transitions \cite{baik2005phase,benaych2012singular}. When prior information about the signal is available, subsequent work has developed approximate message passing (AMP) algorithms and established information-theoretic limits for Bayesian estimation \cite{rangan2012iterative,deshpande2014information,lelarge2019fundamental,montanari2021estimation,barbier2023fundamental,dudeja2026optimality,chen2026orthogonal,chen2026information}. The extensive-rank regime, where the rank grows proportionally with the matrix dimensions, has also received considerable attention, particularly in matrix factorization and dictionary learning \cite{kabashima2016phase,maillard2022perturbative,barbier2022statistical,pourkamali2023bayesian}. In this context, information-theoretic limits and phase transitions have been characterized for Gram matrices with non-Gaussian i.i.d. factors under Gaussian noise \cite{barbier2025phase}.

While matrix factorization aims to recover latent factors,
matrix denoising focuses on estimating their product.
For extensive-rank denoising, rotationally invariant estimators (RIEs) apply nonlinear shrinkage to the observed eigenvalues or singular values while keeping the eigenvectors or singular vectors unchanged. Bun et al. \cite{bun2016rotational} developed RIEs for symmetric
matrices under additive and multiplicative rotationally invariant (RI) perturbations.
For rectangular RI signals under Gaussian noise,
Troiani et al. \cite{troiani2022optimal} derived an asymptotically optimal RIE and characterized its asymptotic MSE.
Related rectangular RIEs under general bi-rotationally invariant
additive noise were developed in
\cite{landau2023singular,pourkamali2025rectangular}.
However, this Gaussian-noise optimality does not generally
extend to i.i.d. non-Gaussian noise.
Indeed, \textit{despite spectral universality, optimal denoising can depend on the details of the noise distribution} \cite[Sec.~V-B]{semerjian2024matrix}.
This motivates efficient algorithms that exploit the noise prior beyond spectral estimation, together with tractable characterizations of their performance.
Related message-passing algorithms have incorporated RIEs for matrix sensing \cite{maillard2024bayes,erba2025bilinear}.
The most closely related work \cite{he2024improved} combines a spectral denoiser based on sub-optimal shrinkage with an entrywise denoiser for compressive robust PCA.
Here, we study this construction for direct extensive-rank matrix denoising without a sensing operator and seek a precise asymptotic characterization of the resulting iteration.

\emph{Contributions.}
We apply the orthogonal approximate message passing (OAMP) framework \cite{ma2014turbo,ma2017orthogonal,takeuchi2020rigorous,he2024improved} to direct extensive-rank matrix denoising using the optimal Gaussian-channel RIE and an entrywise posterior-mean denoiser.
We formulate a Gaussian equivalence conjecture at each fixed iteration: the entrywise module admits a scalar Gaussian channel, while the spectral module matches a Gaussian reference of equal noise variance in input spectrum and RIE MSE.
Under this conjecture, we derive state evolution (SE) and establish a necessary and sufficient condition for its convergence to zero error.
Numerical results support these predictions and demonstrate a larger recovery region than several existing robust-PCA methods.

Detailed proofs, implementation details, and additional numerical results are provided in the appendices.

\section{Model and Assumptions}\label{Sec:model}

We consider the rectangular matrix denoising model
\begin{equation}
\bY=\bL+\bS,
\end{equation}
where $\bY,\bL,\bS\in\R^{m\times n}$.
We take $m \le n$ without loss of generality and assume that $m/n \to \beta \in (0,1]$ as $m, n \to \infty$.
The matrices $\bL$ and $\bS$ are independent. The matrix $\bL$ is bi-rotationally invariant (RI), meaning that its distribution is invariant under left and right multiplication by arbitrary deterministic orthogonal matrices.
The entries of $\bS$ are i.i.d.\ from a (generally non-Gaussian) distribution $P_{\sS}$ with zero mean and variance $p_{\sS} := \E[S_{ij}^{2}] \in (0,\infty)$.

With $O(1)$ entries, $\bL$ and $\bS$ have singular values of
order $\sqrt{n}$. Define $\nu_{L,n}:=m^{-1}\sum_{i=1}^{m}\delta_{\lambda_i(\bL\bL^\top/n)}$.
We assume that $\nu_{L,n}$ converges weakly almost surely
to a deterministic probability measure $\nu_L$ on
$\R_+$ and that
$\limsup_{n\to\infty}\|\bL\|_{\mathrm{op}}/\sqrt n<\infty$
almost surely. The latter gives uniformly bounded
supports, which implies
\[
\frac1{mn}\|\bL\|_{\sF}^2
=\int x\,\nu_{L,n}(\dd x)
\xrightarrow{\mathrm{a.s.}}
\int x\,\nu_L(\dd x)
=:p_{\sL}.
\]

Our goal is to estimate $\bL$ and $\bS$ from $\bY$ given the entrywise prior $P_{\sS}$ and the limiting spectral law $\nu_L$.

\section{OAMP Algorithm and State Evolution}\label{Sec:Framework}

We consider a two-module OAMP iteration following the orthogonalization principle~\cite{ma2014turbo,ma2017orthogonal,takeuchi2020rigorous,rangan2019vector} and the extrinsic construction in~\cite{he2024improved}.
It alternates between entrywise denoising of $\bS$ and spectral denoising of $\bL$ directly from $\bY=\bL+\bS$, without a sensing matrix.
Under the Gaussian equivalence conjecture below, we characterize its asymptotic MSE through a state evolution (SE) recursion.

\subsection{Two-Module OAMP Iteration}\label{Sec:OAMP_algorithm}

Starting from $\bL^{1}:=\bm{0}$, the OAMP iteration updates the entrywise and spectral modules sequentially for $t\in\{1,\ldots,T\}$:
\begin{alignat}{2}
\bR_{\sS}^{t}&=\bY-\bL^{t},\;&\bS^{t}&=\frac{\hat{F}_{\sS}(\bR_{\sS}^{t})-a_{\sS}^{t}\bR_{\sS}^{t}}{1-a_{\sS}^{t}},\label{Eq:Algorithm_S_Update}\\
\bR_{\sL}^{t}&=\bY-\bS^{t},\;&\bL^{t+1}&=\frac{\hat{F}_{\sL}(\bR_{\sL}^{t})-a_{\sL}^{t}\bR_{\sL}^{t}}{1-a_{\sL}^{t}}.\label{Eq:Algorithm_L_Update}
\end{alignat}
Here, $\hat F_{\sS}:\R\to\R$ acts entrywise, whereas
$\hat F_{\sL}:\R^{m\times n}\to\R^{m\times n}$ applies a
scalar shrinkage function $f_{\sL}$ to the singular values
of $\bm R$: given a thin SVD
$\bm R=\bm U\operatorname{diag}(\sigma_1,\ldots,\sigma_m)\bm V^\top$,
\[
\hat F_{\sL}(\bm R)
=
\bm U\operatorname{diag}\!\bigl(
f_{\sL}(\sigma_1),\ldots,f_{\sL}(\sigma_m)
\bigr)\bm V^\top.
\]
After $T$ iterations, the algorithm outputs the denoised estimates $\hat{\bS}^T:=\hat F_{\sS}(\bR_{\sS}^T)$ and $\hat{\bL}^{T+1}:=\hat F_{\sL}(\bR_{\sL}^T)$.

Next, we define $\hat F_{\sS},\hat F_{\sL}$ and $a_{\sS}^t,a_{\sL}^t$
via auxiliary Gaussian-noise models, choosing the scalar MMSE
estimator for $P_{\sS}$ and the optimal RIE \cite{troiani2022optimal}
for $\nu_L$. As formalized in Section~\ref{Sec:SE}, these
auxiliary models capture the conjectured asymptotic behavior
of the module inputs $\bR_{\sS}^t$ and $\bR_{\sL}^t$.

\subsubsection{Entrywise MMSE Estimator}\label{Sec:entrywise}

For $\tau>0$, define the auxiliary Gaussian observation
$\mathsf R_\tau:=\mathsf S+\sqrt{\tau}\,\mathsf Z$ with
independent $\mathsf S\sim P_{\sS}$ and
$\mathsf Z\sim\mathcal N(0,1)$.
The scalar MMSE estimator and its MSE are
\begin{align}
\hat F_{\sS}(r,\tau)
&:=
\E[\mathsf S\mid \mathsf R_\tau=r],
\\
\MMSE_{\sS}(\tau)
&:=
\E\!\left[
\bigl(\mathsf S-\hat F_{\sS}(\mathsf R_\tau,\tau)\bigr)^2
\right].\label{Eqn:mmse_S_def}
\end{align}
Define $a_{\sS}(\tau):=\MMSE_{\sS}(\tau)/\tau$.
At iteration $t$, with $\tau_{\sL}^t$ given by Section~\ref{Sec:SE},
we write $\hat F_{\sS}(\bR_{\sS}^t):=\hat F_{\sS}(\bR_{\sS}^t,\tau_{\sL}^t)$
(applied entrywise) and set $a_{\sS}^t:=a_{\sS}(\tau_{\sL}^t)$.

\subsubsection{Spectral RIE}\label{Sec:RIE}

For $\tau>0$, define the auxiliary Gaussian observation
$\bm{\mathsf R}_\tau:=\bL+\sqrt\tau\,\bm{\mathsf G}$, where
$\bm{\mathsf G}\in\R^{m\times n}$ has i.i.d.\ $\mathcal N(0,1)$
entries and is independent of $\bL$.
Let $\mu_\tau$ denote the limiting empirical distribution of
the normalized singular values
$\sigma_i(\bm{\mathsf R}_\tau)/\sqrt{np_{\sL}}$, $i\in[m]$.
Assuming that $\mu_\tau$ admits a density $\rho_\tau$ on $\R_+$,
its symmetrization $\widehat\mu_\tau$ has density
$\widehat\rho_\tau(x):=\tfrac12\rho_\tau(|x|)$, $x\in\R$, with
unnormalized Hilbert transform
\[
\mathcal H_\tau(s)
:=
\operatorname{p.v.}\!\int_{\R}
\frac{\widehat\rho_\tau(x)}{s-x}\,\dd x,
\qquad s>0,
\]
where $\operatorname{p.v.}$ denotes the Cauchy principal value.
In our notation, the optimal shrinkage function of \cite[Result~1]{troiani2022optimal} is given by
\begin{equation}
\label{Eq:Optimal_Shrinker}
\hat f_{\beta,\tau}(s)
:=
s-\frac{\tau(1-\beta)}{p_{\sL}s}
-\frac{2\beta\tau}{p_{\sL}}\,\mathcal H_\tau(s).
\end{equation}
Given $\bm R\in\R^{m\times n}$ with thin SVD
$\bm U\operatorname{diag}(\sigma_1,\ldots,\sigma_m)\bm V^\top$
and normalized singular values $s_i:=\sigma_i/\sqrt{np_{\sL}}$, define
\begin{equation}
\begin{aligned}[b]
&\operatorname{RIE}_{\beta,\tau}(\bm R)\\
&\quad:=\sqrt{np_{\sL}}\,\bm U\operatorname{diag}\!\bigl(
\hat f_{\beta,\tau}(s_1),\ldots,\hat f_{\beta,\tau}(s_m)
\bigr)\bm V^\top.
\end{aligned}
\end{equation}

For this Gaussian channel, the RIE is asymptotically MMSE-optimal,
with MSE given by \cite[Results~1 and~2]{troiani2022optimal}
\begin{equation}\label{Eqn:mmse_L_def}
\begin{aligned}[b]
&\MMSE_{\sL}(\tau):=\lim_{n\to\infty}\frac1{mn}
\E\left\|\bL
-\operatorname{RIE}_{\beta,\tau}(\bm{\mathsf R}_\tau)
\right\|_{\sF}^{2}\\
&=\tau-\frac{\tau^2}{p_{\sL}}
\int_{\R_+}\!\left[\frac{(1-\beta)^2}{s^2}
+\frac{\pi^2\beta^2\rho_\tau(s)^2}{3}\right]\mu_\tau(\dd s).
\end{aligned}
\end{equation}
Define the spectral MSE ratio $a_{\sL}(\tau):=\MMSE_{\sL}(\tau)/\tau$
for $\tau>0$, whose analytical properties are summarized next.

\begin{lemma}[Spectral MSE ratio]
\label{Lemma:Spectral_Ratio}
The function $a_{\sL}$ is continuous and strictly decreasing on
$(0,\infty)$, with $0<a_{\sL}(\tau)\le p_{\sL}/(p_{\sL}+\tau)<1$
and $\lim_{\tau\to\infty}a_{\sL}(\tau)=0$.
Moreover, $d_{\sL}:=\lim_{\tau\downarrow0}a_{\sL}(\tau)$ exists in
$(0,1]$; setting $a_{\sL}(0):=d_{\sL}$ extends $a_{\sL}$ to a
continuous, strictly decreasing function on $[0,\infty)$.
\end{lemma}

A proof of Lemma~\ref{Lemma:Spectral_Ratio} is provided in Appendix~\ref{Appendix:Rectangular_Risk}. At iteration $t$, with $\tau_{\sS}^t$ specified in Section~\ref{Sec:SE},
we set $\hat F_{\sL}(\bR_{\sL}^t):=
\operatorname{RIE}_{\beta,\tau_{\sS}^t}(\bR_{\sL}^t)$ and
$a_{\sL}^t:=a_{\sL}(\tau_{\sS}^t)$.

\subsection{Orthogonality Property}

The updates in \eqref{Eq:Algorithm_S_Update}--\eqref{Eq:Algorithm_L_Update} follow the construction in \cite{ma2017orthogonal,he2024improved}, which subtract scaled inputs to make the update \textit{divergence-free} (or extrinsic). For the entrywise posterior-mean denoiser, the corrected output error is orthogonal to the input noise on its auxiliary Gaussian channel, as in standard OAMP \cite{ma2017orthogonal,takeuchi2020rigorous}.

For the spectral module, set
$\hat F_{\sL}:=\operatorname{RIE}_{\beta,\tau}$ on the Gaussian channel
$\bm{\mathsf R}_\tau=\bL+\sqrt{\tau}\,\bm{\mathsf G}$, and define
\[
\bm{\mathsf L}_\tau:=
\frac{\hat F_{\sL}(\bm{\mathsf R}_\tau)
-a_{\sL}(\tau)\bm{\mathsf R}_\tau}{1-a_{\sL}(\tau)}.
\]
Then (see Appendix~\ref{Appendix:Calibration}),
\begin{equation}
\label{Eq:Spectral_Orthogonality}
\lim_{n\to\infty}
\frac{1}{mn}\,\E\langle\bm{\mathsf L}_\tau,\bm{\mathsf G}\rangle_{\sF}=0,
\end{equation}
where $\langle\bm A,\bm B\rangle_{\sF}:=\operatorname{tr}(\bm A^\top\bm B)$.
Since $\bL$ is independent of $\bm{\mathsf G}$, the output error $\bm{\mathsf L}_\tau-\bL$ is likewise asymptotically orthogonal to the input noise $\bm{\mathsf R}_\tau-\bL=\sqrt{\tau}\,\bm{\mathsf G}$.

\subsection{State Evolution and Gaussian Equivalence Conjecture}\label{Sec:SE}

Starting from $\tau_{\sL}^{1}:=p_{\sL}$, define the
state evolution (SE) sequence by
\begin{equation}
\label{Eq:SE_Recursion}
\tau_{\sS}^{t}
=\Phi_{\sS}(\tau_{\sL}^{t}),
\qquad
\tau_{\sL}^{t+1}
=\Phi_{\sL}(\tau_{\sS}^{t}),
\end{equation}
where
$\Phi_\bullet(v):=
v\MMSE_\bullet(v)/(v-\MMSE_\bullet(v))$
for $\bullet\in\{\sS,\sL\}$, and $\MMSE_{\sS}$ and $\MMSE_{\sL}$
are given by \eqref{Eqn:mmse_S_def} and \eqref{Eqn:mmse_L_def}.

\begin{conjecture}[Gaussian equivalence]
\label{Conj:Effective}
Let $(\tau_{\sS}^{t},\tau_{\sL}^{t+1})_{t\ge1}$ be given by
\eqref{Eq:SE_Recursion}, and denote by $W_2$ the $2$-Wasserstein
distance. For each fixed $t\ge1$ with $\tau_{\sL}^{t},\tau_{\sS}^{t}>0$,
under suitable regularity conditions on the signal model, the
following hold in probability as $m,n\to\infty$ with $m/n\to\beta$:

\begin{enumerate}
\item[\textup{(i)}]
With $\mathsf S$ and $\mathsf R_\tau$ defined in
Section~\ref{Sec:OAMP_algorithm}, let
$\mathsf R_{\sS}^{t}:=\mathsf R_{\tau_{\sL}^{t}}$ and
$\mathsf S^{t}:=(1-a_{\sS}^{t})^{-1}
\bigl(\hat F_{\sS}(\mathsf R_{\sS}^{t})
-a_{\sS}^{t}\mathsf R_{\sS}^{t}\bigr)$,
and denote their joint law with $\mathsf S$ by
$\mathcal L(\mathsf S,\mathsf R_{\sS}^{t},\mathsf S^{t})$. Then
\begin{equation}
\label{Eq:Entrywise_Equivalence}
W_2\!\left(\frac{1}{mn}\sum_{i,j}
\delta_{(S_{ij},R_{\sS,ij}^{t},S_{ij}^{t})},\,
\mathcal L(\mathsf S,\mathsf R_{\sS}^{t},\mathsf S^{t})\right)\to0.
\end{equation}

\item[\textup{(ii)}]
For $\bm M\in\R^{m\times n}$, let
$\mu_n(\bm M):=m^{-1}\sum_{i=1}^{m}
\delta_{\sigma_i(\bm M)/\sqrt n}$
be the empirical distribution of its normalized singular values,
where $\sigma_1(\bm M)\ge\cdots\ge\sigma_m(\bm M)$
(assuming $m\le n$).
With $\bm{\mathsf R}_{\tau}$ defined in Section~\ref{Sec:RIE},
\begin{equation}
\label{Eq:Spectral_ESD_Equivalence}
W_2\bigl(
\mu_n(\bR_{\sL}^{t}),\,
\mu_n(\bm{\mathsf R}_{\tau_{\sS}^{t}})
\bigr)
\to0,
\end{equation}
and the denoiser MSE satisfies
\begin{equation}
\label{Eq:Spectral_Risk_Equivalence}
\frac{1}{mn}
\bigl\|
\bL-\hat F_{\sL}(\bR_{\sL}^{t})
\bigr\|_{\sF}^{2}
\to
\MMSE_{\sL}(\tau_{\sS}^{t}).
\end{equation}
\end{enumerate}
\end{conjecture}

\begin{remark}
(i) The Gaussian conditioning arguments used to establish AMP state evolution \cite{bayati2011dynamics} are not directly applicable here. (ii) We state SE only for the Bayes-optimal choices of the entrywise and spectral denoisers, which admit simplifications (e.g., no constant matrix in the input of the spectral denoiser). Numerical results suggest that an analogous SE holds more generally, but we do not include it in this paper.
\end{remark}

\section{Recovery Condition via State Evolution}
\label{Sec:Simultaneous_Recovery}

Under Conjecture~\ref{Conj:Effective}, exact asymptotic recovery of
$\bL$ and $\bS$ corresponds to $\tau_{\sS}^{t},\tau_{\sL}^{t}\to 0$
as $t\to\infty$ in \eqref{Eq:SE_Recursion}. Whether this holds depends
on the limiting spectral law $\nu_L$ and the entrywise prior $P_{\sS}$.
With $p_{\sL}>0$ and $p_{\sS}>0$ defined in Section~\ref{Sec:model},
we factor out the power scales by writing
$\bL=\sqrt{p_{\sL}}\,\widetilde{\bL}$ and
$\bS=\sqrt{p_{\sS}}\,\widetilde{\bS}$, so that the entries of
$\widetilde{\bS}$ follow a zero-mean, unit-variance prior
$\widetilde P_{\sS}$ and $\widetilde{\bL}\widetilde{\bL}^\top/n$ has
limiting spectral law $\widetilde\nu_L$ satisfying
$\int x\,\widetilde\nu_L(\dd x)=1$. Once the SE variables are normalized by $p_{\sL}$, exact recovery
depends on $(p_{\sL},p_{\sS})$ through the power ratio
$\zeta:=p_{\sS}/p_{\sL}>0$, together with the structural parameters
(such as rank and sparsity) encoded in $\widetilde\nu_L$ and
$\widetilde P_{\sS}$.

We assume that neither $\widetilde P_{\sS}$ nor $\widetilde\nu_L$ has a
singular-continuous component. Specifically, we decompose the
unit-variance entrywise prior as
\begin{equation}\label{Eqn:Ps_mix}
\widetilde P_{\sS}=(1-\rho_{\sS})P_{\sS,\mathrm d}
+\rho_{\sS}P_{\sS,\mathrm c},
\end{equation}
where $P_{\sS,\mathrm d}$ and $P_{\sS,\mathrm c}$ are discrete and
absolutely continuous probability measures, respectively, and
$\rho_{\sS}\in[0,1]$. Similarly, we decompose the normalized limiting
spectral law on $\R_+$ as
\begin{equation}\label{Eqn:PL_mix}
\widetilde\nu_L=w_0\delta_0
+\sum_{\xi\in\mathcal A}w_\xi\delta_\xi
+\nu_{L,\mathrm c},
\end{equation}
where $w_0:=\widetilde\nu_L(\{0\})$,
$\mathcal A:=\{\xi>0:\widetilde\nu_L(\{\xi\})>0\}$ is the set of
positive atoms with weights $w_\xi:=\widetilde\nu_L(\{\xi\})$, and
$\nu_{L,\mathrm c}$ is the absolutely continuous part of
$\widetilde\nu_L$.

For $\bullet\in\{\sS,\sL\}$ and $x>0$, define
\begin{equation}\label{Eq:Normalized_Transfers}
\begin{aligned}
\bar{\MMSE}_{\bullet}(x)&:=\frac{\MMSE_{\bullet}(xp_{\bullet})}{p_{\bullet}},\quad
\bar a_{\bullet}(x):=\frac{\bar{\MMSE}_{\bullet}(x)}x,\\
\bar\Psi_{\bullet}(x)&:=\frac{x\bar a_{\bullet}(x)}{1-\bar a_{\bullet}(x)}=\frac{\Phi_\bullet(p_\bullet x)}{p_\bullet}.
\end{aligned}
\end{equation}
Equivalently, $\bar{\MMSE}_{\bullet}(x)$, $\bar a_{\bullet}(x)$, and
$\bar\Psi_{\bullet}(x)$ are the counterparts of $\MMSE_{\bullet}$,
$a_{\bullet}$, and $\Phi_{\bullet}$ for the unit-power model
($\widetilde P_{\sS}$ or $\widetilde\nu_L$) at input noise variance
$x$.

\begin{lemma}[Properties of the normalized MSE functions]
\label{Lemma:Risk_Conditions}
For each $\bullet\in\{\sL,\sS\}$, the function $\bar\Psi_\bullet$
is continuous and nondecreasing on $(0,\infty)$, with
$0<\bar\Psi_\bullet(x)\le1$ for $x>0$.
Moreover, $\bar a_\bullet$ is continuous on $(0,\infty)$,
satisfies $0<\bar a_\bullet(x)\le(1+x)^{-1}$ for $x>0$,
and tends to zero as $x\to\infty$.
If $d_\bullet:=\lim_{x\downarrow0}\bar a_\bullet(x)$ exists
and satisfies $d_\bullet<1$, then
$\bar\Psi_\bullet(x)=\tfrac{d_\bullet}{1-d_\bullet}x+o(x)$
as $x\downarrow0$.
In this case, $\bar\Psi_\bullet$ admits a continuous extension
to the origin, with $\bar\Psi_\bullet(0)=0$.
\end{lemma}

The quantity $d_{\sS}$ is the MMSE dimension of $\widetilde P_{\sS}$
\cite{Wu2011MMSE}, and $d_{\sL}$ is its spectral counterpart, which we
term the \emph{spectral MMSE dimension} of $\widetilde\nu_L$.
Define $D:=d_{\sL}+d_{\sS}$ and
$\Delta_{\sS}:=\sup_{x>0}\bar a_{\sS}(x)-d_{\sS}$.

\begin{theorem}[SE recovery regimes]
\label{Theorem:Power_Regimes}
Fix $\beta\in(0,1]$, and let $\widetilde\nu_L$ and $\widetilde P_{\sS}$
satisfy the assumptions of Lemmas~\ref{Lemma:Spectral_Ratio}
and~\ref{Lemma:Risk_Conditions}, with $0<d_{\sL}<1$, $0\le d_{\sS}<1$,
and $D\ne 1$ whenever $\Delta_{\sS}>0$. For the SE recursion
\eqref{Eq:SE_Recursion} initialized at $\tau_{\sL}^{1}=p_{\sL}$, we
have $(\tau_{\sS}^{t},\tau_{\sL}^{t})\to(0,0)$ as $t\to\infty$ if and
only if $\zeta>\zeta_{\sC}$, where
\begin{equation}
\label{Eq:Zeta_Critical}
\zeta_{\sC}
:=\sup_{x>0}
\frac{\bar a_{\sL}^{-1}\bigl(1-\bar a_{\sS}(x)\bigr)}
     {\bar\Psi_{\sS}(x)}
\in[0,\infty],
\end{equation}
with $\bar a_{\sL}^{-1}:[0,1]\to[0,\infty]$ extending the inverse of
$\bar a_{\sL}$ on $(0,d_{\sL})$ via $\bar a_{\sL}^{-1}(y):=0$ for
$y\in[d_{\sL},1]$ and $\bar a_{\sL}^{-1}(0):=\infty$. Further,
three regimes arise:
\begin{enumerate}
\item \emph{No recovery} ($D>1$): $\zeta_{\sC}=\infty$, so
$(\tau_{\sS}^{t},\tau_{\sL}^{t})$ fails to converge to $(0,0)$ for
any $\zeta>0$; moreover, $(0,0)$ is locally unstable.

\item \emph{Power-dependent recovery} ($D<1<D+\Delta_{\sS}$):
$\zeta_{\sC}\in(0,\infty)$ with the supremum in
\eqref{Eq:Zeta_Critical} attained. Here $(0,0)$ is locally stable for
every $\zeta>0$, whereas $(\tau_{\sS}^{t},\tau_{\sL}^{t})$ converges to
a nonzero fixed point whenever $0<\zeta\le\zeta_{\sC}$.

\item \emph{All-power recovery} ($D+\Delta_{\sS}\le 1$):
$\zeta_{\sC}=0$, so $(\tau_{\sS}^{t},\tau_{\sL}^{t})\to(0,0)$ for
every $\zeta>0$.
\end{enumerate}
\end{theorem}

The proofs of Lemma~\ref{Lemma:Risk_Conditions} and Theorem~\ref{Theorem:Power_Regimes} are provided in Appendices~\ref{Appendix:Extrinsic_Monotonicity} and~\ref{Appendix:Simultaneous_Recovery}, respectively.

\begin{remark}
In the power-dependent regime,
$\zeta=p_{\sS}/p_{\sL}>\zeta_{\sC}$ shows that
increasing the relative power of non-Gaussian noise
$\bS$ can enable exact recovery.
This may seem counterintuitive, but similar phenomena have been
observed in scalar non-Gaussian channels, where higher noise power can
reduce the MMSE \cite{Wu2011MMSE}.
\end{remark}

\section{Numerical Results}\label{Sec:Numerical}

We evaluate the proposed algorithm on robust PCA, where $\bL$ is low-rank and $\bS$ is sparse. We generate $\bL=\sqrt{p_{\sL}/r}\,\bA\bB$, where $\bA\in\R^{m\times r}$ and $\bB\in\R^{r\times n}$ have i.i.d.\ standard Gaussian entries, and $\gamma:=r/m$ denotes the normalized rank. The entries of $\bS$ are drawn independently from the Bernoulli--Gaussian prior $S_{ij}\sim(1-\rho)\delta_0+\rho\Normal(0,p_{\sS}/\rho)$, where $\rho\in(0,1)$ denotes the sparsity ratio. In our experiments, the entrywise denoiser MSE is estimated by averaging the coordinate-wise posterior variances, while the spectral denoiser MSE is estimated using Stein's unbiased risk estimate (SURE)~\cite{he2024improved}.

\begin{figure}[!t]
\centering
\setlength{\abovecaptionskip}{2pt}
\includegraphics[width=\columnwidth]{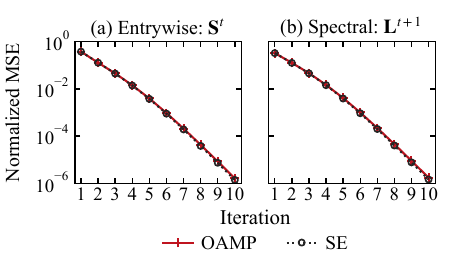}
\caption{Empirical normalized MSEs of OAMP versus SE predictions.}
\label{Fig:Main_RPCA_Multi}
\end{figure}

\textit{Numerical Validation of the Conjecture.}
In Fig.~\ref{Fig:Main_RPCA_Multi}, we plot the normalized MSEs
$e_{\sS}^{t}:=\|\bS-\bS^{t}\|_{\sF}^{2}/(mn p_{\sS})$
and
$e_{\sL}^{t+1}:=\|\bL-\bL^{t+1}\|_{\sF}^{2}/(mn p_{\sL})$,
along with their state evolution (SE) predictions.
We set $m=n=2000$ ($\beta=1$), $r=600$ ($\gamma=0.3$), $\rho=0.15$, and
$p_{\sL}=p_{\sS}=1$, initialize the algorithm at
$\bL^{1}=\bm{0}$ and $\widehat{\tau}_{\sL}^{1}=p_{\sL}$, and average the
empirical MSEs over 10 independent trials.
As shown in the figure, the SE predictions closely track the empirical MSEs across
iterations.
Additional numerical tests confirm close agreement between the empirical effective-noise distributions and their Gaussian references, further supporting Conjecture \ref{Conj:Effective}.

\begin{figure}[!t]
\centering
\setlength{\abovecaptionskip}{2pt}
\includegraphics[width=\columnwidth]{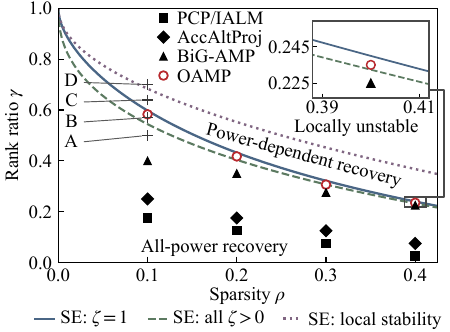}
\caption{Recovery phase diagram ($\zeta=1$, $\beta=1$).}
\label{Fig:Recovery_Phase_Diagram_1}
\end{figure}

\textit{Recovery Phase Diagram.} Fig.~\ref{Fig:Recovery_Phase_Diagram_1} plots the empirical recovery phase diagram in the $(\rho,\gamma)$-plane. In this experiment, we set $m=n=2000$ and $p_{\sL}=p_{\sS}=1$ ($\zeta=1$, $\beta=1$). For the signal priors used in this experiment, $d_{\sS}=\rho$ and $d_{\sL}=\gamma(1+\beta-\beta\gamma)$. Notice that Theorem~\ref{Theorem:Power_Regimes} characterizes recovery via $\zeta>\zeta_{\sC}$ for fixed normalized priors. For a fixed $\zeta=1$, this condition can be translated into regions of $(\rho,\gamma)$ through the dependence $\zeta_{\sC}=\zeta_{\sC}(\rho,\gamma)$. Specifically, the dashed curve ($2\gamma-\gamma^2+\rho+\Delta_{\sS}=1$) marks the all-power recovery boundary below which $\zeta_{\sC}(\rho,\gamma)=0$, and the dotted curve ($2\gamma-\gamma^2+\rho=1$) marks the local-stability boundary above which $\zeta_{\sC}(\rho,\gamma)=\infty$. Between these curves, recovery is power-dependent, and the condition $\zeta_{\sC}(\rho,\gamma)=1$ defines the solid curve separating success and failure at $\zeta=1$.

In Fig.~\ref{Fig:Recovery_Phase_Diagram_1}, we also compare OAMP with
principal component pursuit via inexact augmented Lagrange multipliers
(PCP/IALM)~\cite{candes2011robust,lin2010alm}, accelerated alternating
projections (AccAltProj)~\cite{cai2019accaltproj}, and
BiG-AMP~\cite{parker2014bigamp2}.
OAMP achieves a larger recovery region than all three baseline methods.

\begin{figure}[!t]
\centering
\setlength{\abovecaptionskip}{2pt}
\includegraphics[width=\columnwidth]{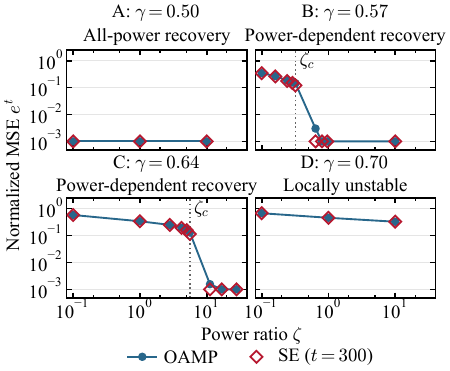}
\caption{Maximum of the two normalized extrinsic MSEs $e^t$ versus $\zeta$ at points A--D ($\rho=0.1$; values below $10^{-3}$ clipped to $10^{-3}$).}
\label{Fig:Recovery_Phase_Diagram_2}
\end{figure}

In Fig.~\ref{Fig:Recovery_Phase_Diagram_2}, we plot the reconstruction
MSEs across varying $\zeta$ for four $(\rho,\gamma)$ configurations
(marked A, B, C, and D in Fig.~\ref{Fig:Recovery_Phase_Diagram_1}).
OAMP recovers the signals at all tested $\zeta$ for A, at sufficiently
large $\zeta$ for B and C, and at no tested $\zeta$ for D, consistent with
Theorem~\ref{Theorem:Power_Regimes}.

\section{Conclusion}

We studied an OAMP iteration for direct extensive-rank matrix denoising under i.i.d. non-Gaussian noise and formulated a Gaussian equivalence conjecture for its asymptotic behavior. The associated SE yields convergence and recovery conditions supported by numerical results. Proving the conjecture and determining whether, and in which regimes, the iteration attains Bayes-optimal performance in the original non-Gaussian denoising model remain open problems.

\clearpage
\makeatletter
\let\IEEEoriginalmakecaption\@makecaption
\long\def\@makecaption#1#2{%
\ifx\@captype\@IEEEtablestring
\IEEEoriginalmakecaption{#1}{#2}%
\else
\@IEEEfigurecaptionsepspace
{\normalfont\footnotesize #1.\quad #2\par}%
\fi}
\makeatother

\makeatletter
\setlength{\@fptop}{0pt}
\setlength{\@dblfptop}{0pt}
\makeatother
\appendices
\allowdisplaybreaks[0]

\section{Derivations of the Denoisers and MSE}
\label{Appendix:Calibration}

This appendix collects the derivations of the two
denoisers and the resulting MSE transfer functions. It also describes
the implementation for Figs.~\ref{Fig:Main_RPCA_Multi} and~\ref{Fig:Recovery_Phase_Diagram_1}
in Section~\ref{Sec:Numerical} and presents the effective-noise and
input-spectrum comparisons in Figs.~\ref{Fig:Check_Effective_Noise}--\ref{Fig:Check_Observed_Input_Spectrum}.

\subsection{Spectral Notation and Extension}
\label{Appendix:Reference_Conditions}

We use $\mu_\tau$, $\widehat\mu_\tau$, $\rho_\tau$,
$\mathcal H_\tau$, and $\hat f_{\beta,\tau}$ from
Section~\ref{Sec:Framework}.
Let $\nu$ be the law of $\sqrt{x}$ when
$x\sim\widetilde\nu_{L}$, and let $\widehat\nu$ be its
symmetrization. Thus $\nu$ is supported on $[0,M]$ for some
$M<\infty$ and $\int s^2\nu(\dd s)=1$.
On $\operatorname{supp}\mu_\tau\cap(0,\infty)$,
\eqref{Eq:Optimal_Shrinker} reads
\begin{equation}
\hat f_{\beta,\tau}(s)
=s-\frac{\tau}{p_{\sL}}\left[
\frac{1-\beta}{s}+2\beta\mathcal H_\tau(s)\right].
\label{Eq:Appendix_RIE_Profile}
\end{equation}

For each $\tau>0$, the boundary-value results in
\cite[Thm.~2.1]{dozier2007analysis},
\cite[Prop.~2.3]{capitaine2014separation}, and
\cite[Cor.~1]{biane1997semicircular} give continuity on the
positive spectral support and the value $\hat f_{\beta,\tau}(0)=0$ when the
support reaches zero. Together with
Lemma~\ref{Lemma:Noisy_Spectrum_RIE}, they give a bounded,
continuous, nondecreasing extension to $[0,\infty)$.

\subsection{Correction and Scaling Coefficients}
\label{Appendix:Extrinsic_Coefficients}

\subsubsection{General denoisers}
For a denoiser $F$ on the auxiliary channel
$\bm{\mathsf R}_\tau=\bL+\sqrt{\tau}\,\bm{\mathsf G}$,
we form the divergence-free estimate
\cite{he2024improved}
\begin{equation*}
\widetilde F(\bm{\mathsf R}_\tau)
=c_F\bigl[F(\bm{\mathsf R}_\tau)-a_F\bm{\mathsf R}_\tau\bigr].
\end{equation*}
We choose $a_F$ so that
$\E\langle F(\bm{\mathsf R}_\tau)-a_F\bm{\mathsf R}_\tau,
\bm{\mathsf G}\rangle_{\sF}=0$.
Since $\E\langle\bm{\mathsf R}_\tau,\bm{\mathsf G}\rangle_{\sF}
=mn\sqrt{\tau}$, this gives
\begin{equation}
a_F=\frac{\E\langle F(\bm{\mathsf R}_\tau),
\bm{\mathsf G}\rangle_{\sF}}{mn\sqrt{\tau}}.
\label{Eq:ITMP_Decorrelation_Parameter}
\end{equation}

Fixing $a_F$, we minimize the quadratic MSE
$\E\|\widetilde F(\bm{\mathsf R}_\tau)-\bL\|_{\sF}^2$
over $c_F$ to obtain
\begin{equation}
c_F^\star=
\frac{\E\langle F(\bm{\mathsf R}_\tau)-a_F\bm{\mathsf R}_\tau,
\bm{\mathsf R}_\tau\rangle_{\sF}}
{\E\|F(\bm{\mathsf R}_\tau)-a_F\bm{\mathsf R}_\tau\|_{\sF}^2}.
\label{Eq:General_Extrinsic_Scaling}
\end{equation}

\subsubsection{Posterior-mean denoisers}
Suppose $F(\bm{\mathsf R}_\tau)=\E[\bL\mid\bm{\mathsf R}_\tau]$
and let
\begin{equation*}
\varepsilon:=\frac1{mn}
\E\|F(\bm{\mathsf R}_\tau)-\bL\|_{\sF}^2.
\end{equation*}
Orthogonality of the posterior-mean estimator gives
\begin{equation*}
\E\langle F(\bm{\mathsf R}_\tau),
\sqrt{\tau}\,\bm{\mathsf G}\rangle_{\sF}=mn\varepsilon.
\end{equation*}
Hence $a_F=\varepsilon/\tau$.
Posterior orthogonality and the imposed noise orthogonality also give
\begin{align*}
&\E\|F(\bm{\mathsf R}_\tau)-a_F\bm{\mathsf R}_\tau\|_{\sF}^2=(1-a_F)
\E\langle F(\bm{\mathsf R}_\tau)-a_F\bm{\mathsf R}_\tau,
\bm{\mathsf R}_\tau\rangle_{\sF}.
\end{align*}
An optimal scaling and the resulting MSE per entry are
\begin{equation}
c_F=\frac1{1-a_F},\qquad
\Phi_F(\tau)=\frac{\varepsilon}{1-a_F}
=\frac{\tau\varepsilon}{\tau-\varepsilon}.
\label{Eq:Posterior_Extrinsic_Scaling}
\end{equation}

\subsubsection{Application to the RIE}
Reference~\cite[Result~2]{troiani2022optimal} shows that the RIE
achieves the MMSE asymptotically:
\begin{equation}
\MMSE_{\sL}(\tau)
=p_{\sL}\left(1-\int\hat f_{\beta,\tau}(s)^2\,
\mu_\tau(\dd s)\right).
\label{Eq:Rectangular_Optimal_Energy}
\end{equation}
Also, $\int s^2\,\mu_\tau(\dd s)=1+\tau/p_{\sL}$ and
$\int s\mathcal H_\tau(s)\,\mu_\tau(\dd s)=1/2$.
Thus \eqref{Eq:Appendix_RIE_Profile} gives
\begin{equation}
\int s\hat f_{\beta,\tau}(s)\,\mu_\tau(\dd s)=1.
\label{Eq:Appendix_Spectral_Moments}
\end{equation}
Furthermore,
\begin{align}
\frac{\E\langle\operatorname{RIE}_{\beta,\tau}(\bm{\mathsf R}_{\tau}),
\bm{\mathsf R}_\tau\rangle_{\sF}}{mn}
&\longrightarrow p_{\sL},\nonumber\\
\frac{\E\langle\operatorname{RIE}_{\beta,\tau}(\bm{\mathsf R}_{\tau}),
\bL\rangle_{\sF}}{mn}
&\longrightarrow p_{\sL}-\MMSE_{\sL}(\tau),\nonumber\\
\frac{\E\|\operatorname{RIE}_{\beta,\tau}(\bm{\mathsf R}_\tau)\|_{\sF}^2}{mn}
&\longrightarrow p_{\sL}-\MMSE_{\sL}(\tau).
\label{Eq:Calibration_Optimal_Gram}
\end{align}
These limits satisfy the same moment identities as a posterior mean.
The preceding calculation in \eqref{Eq:Posterior_Extrinsic_Scaling}
therefore gives the correction coefficient
and an optimal scaling for the limiting MSE:
\begin{align}
a_{\sL}(\tau)&=\frac{\MMSE_{\sL}(\tau)}{\tau},
\label{Eq:Calibration_Optimal_Response}
\\
c_{\sL}(\tau)&=\frac1{1-a_{\sL}(\tau)},\qquad
\Phi_{\sL}(\tau)=\frac{\tau\MMSE_{\sL}(\tau)}
{\tau-\MMSE_{\sL}(\tau)}.
\label{Eq:Calibration_Optimal_Risk}
\end{align}
Lemma~\ref{Lemma:Spectral_Ratio} ensures $a_{\sL}<1$.
The defining orthogonality in
\eqref{Eq:ITMP_Decorrelation_Parameter} also gives
\eqref{Eq:Spectral_Orthogonality}.

\subsection{Numerical Implementation}
\label{Appendix:RIE_Regularization}

\subsubsection{Spectral denoising}
We use the known Gaussian-product spectrum from
Section~\ref{Sec:Numerical}.
Its noisy transform satisfies an algebraic equation
\cite[Sec.~3.3 and App.~F]{troiani2022optimal}.
For $\tau>0$, define
\begin{equation*}
g_\tau(z):=\int_{\R}\frac{\widehat\mu_\tau(\dd x)}{z-x}.
\end{equation*}
Denote $d:=1-\beta+\beta z g_\tau(z)$.
In our normalization, $g_\tau$ satisfies
\begin{align}
0&=d\left(z^2-\frac{\tau}{p_{\sL}}d\right)g_\tau(z)^2
+\gamma z^2\nonumber\\
&\quad-z\left[\gamma z^2+
\left(1-\gamma-\frac{\gamma\tau}{p_{\sL}}\right)d\right]g_\tau(z).
\label{Eq:Gaussian_Product_Transform}
\end{align}
We take the analytic branch with $zg_\tau(z)\to1$ at infinity and
$\operatorname{Im}g_\tau(z)<0$ for $\operatorname{Im}z>0$.
{Further}, $\mathcal H_\tau(s)=\operatorname{Re}g_\tau(s+\mathrm i0)$ and
$\rho_\tau(s)=-(2/\pi)\operatorname{Im}g_\tau(s+\mathrm i0)$.

In our implementation, we use
$\eta=10^{-8}\sqrt{\widehat\tau/p_{\sL}}$ and
estimate the optimal shrinkage function as
\begin{align}
\hat f_{\beta,\widehat\tau,\eta}(s)
&=\biggl[s-\frac{\widehat\tau}{p_{\sL}}
\biggl\{(1-\beta)\frac{s}{s^2+\eta^2}\nonumber\\
&\qquad\quad+2\beta\operatorname{Re}g_{\widehat\tau}(s+\mathrm i\eta)
\biggr\}\biggr]_+.
\label{Eq:Regularized_RIE_Profile}
\end{align}
Here $[x]_+=\max\{x,0\}$.
Applying $\hat f_{\beta,\widehat\tau,\eta}$ to the normalized singular values and
rescaling by $\sqrt{np_{\sL}}$ defines $\hat F_{\sL,\eta}$.

For $f=\hat f_{\beta,\widehat\tau,\eta}$ and
$s_i=\sigma_i(\bR)/\sqrt{np_{\sL}}$,
the divergence of the RIE is
\cite{candes2013sure}
\begin{align}
\operatorname{div}\hat F_{\sL,\eta}(\bR)
&=\sum_{i=1}^m\left[f'(s_i)+(n-m)\frac{f(s_i)}{s_i}\right]
\nonumber\\
&\quad+2\sum_{i<j}
\frac{s_i f(s_i)-s_j f(s_j)}{s_i^2-s_j^2}.
\label{Eq:Spectral_Divergence}
\end{align}
The empirical divergence and the MSE estimate via SURE are
\cite{candes2013sure}
\begin{align}
\widehat a_{\mathrm{div},\eta}
&=\frac{\operatorname{div}\hat F_{\sL,\eta}(\bR)}{mn},\nonumber\\
\widehat\MMSE_{\sL,\eta}
&=\frac{\|\hat F_{\sL,\eta}(\bR)-\bR\|_{\sF}^2}{mn}
+2\widehat\tau\widehat a_{\mathrm{div},\eta}-\widehat\tau.
\label{Eq:Empirical_RIE_Coefficients}
\end{align}

\subsubsection{Coefficient and variance updates}
Each module receives $\widehat\tau$ from the preceding module.
Using its MSE estimate $\widehat\MMSE$, we set
\begin{equation*}
a_{\mathrm{used}}=\frac{\widehat\MMSE}{\widehat\tau},\qquad
c_{\mathrm{used}}=\frac1{1-a_{\mathrm{used}}},\qquad
\widehat\tau_{\mathrm{out}}=\frac{\widehat\MMSE}{1-a_{\mathrm{used}}}.
\end{equation*}
These quantities are used in the OAMP updates.

\subsubsection{Details of the phase diagram experiments}
For Fig.~\ref{Fig:Recovery_Phase_Diagram_1}, we use the dimensions
and powers specified in Section~\ref{Sec:Numerical}.
PCP/IALM uses $\lambda=1/\sqrt n$.
AccAltProj uses the true rank and uses threshold decay $0.7$.
BiG-AMP uses the matched Gaussian-factor and Bernoulli--Gaussian
priors and a damping parameter of $0.25$.

For each of the three compared methods and each
$\rho\in\{0.1,0.2,0.3,0.4\}$, we report the largest tested
$\gamma$ for which
$\operatorname{median}_{1\le j\le 10}
 \bigl(\min_{1\le t\le 300}e_j^t\bigr)<10^{-3}$
over 10 independent instances, with $e_j^t$ denoting the
maximum of the two normalized MSEs at iteration $t$ in instance $j$.

At $\rho=0.4$, BiG-AMP recovers all 10 instances at $\gamma=0.225$. We also test $\gamma=0.235$ ($r=470$),
just below the equal-power SE boundary $\gamma\approx0.2399$.
On the same 10 matrix pairs, OAMP recovers both components in every
trial within 64 iterations, whereas BiG-AMP does not achieve
simultaneous recovery in any trial within 300 iterations.

\subsubsection{Gaussian equivalence conjecture}
\begin{figure}[!t]
\centering
\includegraphics[width=\columnwidth]{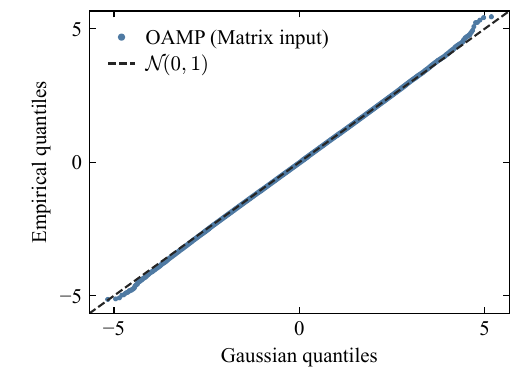}
\caption{Q--Q plot of the normalized entrywise effective noise
at $t=10$ against $\Normal(0,1)$, using all
$mn=4\times10^6$ entries.}
\label{Fig:Check_Effective_Noise}
\end{figure}

\begin{figure}[!t]
\centering
\includegraphics[width=\columnwidth]{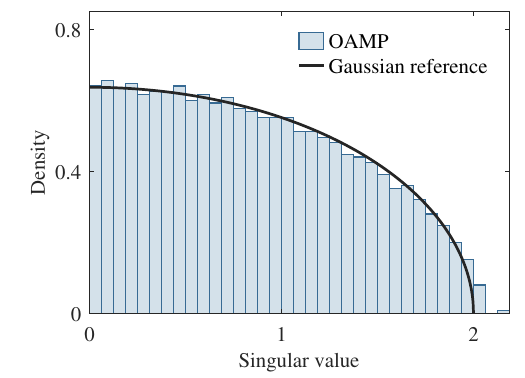}
\caption{Singular-value distributions of
$\widetilde{\bR}_{\sL}^t/\sqrt n$ and the Gaussian reference
$\bm{\mathsf G}/\sqrt n$ at $t=10$.}
\label{Fig:Check_Spectral_Noise}
\end{figure}

\begin{figure}[!t]
\centering
\includegraphics[width=\columnwidth]{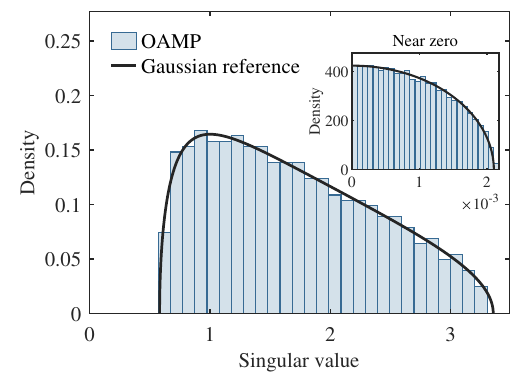}
\caption{Singular-value distribution of
$\bR_{\sL}^t/\sqrt{np_{\sL}}$ and the Gaussian-reference density
$\rho_{\widehat\tau_{\sS}^t}$, for $t=10$.
The inset shows the near-zero spectral band.}
\label{Fig:Check_Observed_Input_Spectrum}
\end{figure}

For the settings and initialization of Fig.~\ref{Fig:Main_RPCA_Multi},
define
\begin{equation*}
\widetilde{\bR}_{\sS}^t
:=\frac{\bR_{\sS}^t-\bS}{\sqrt{\widehat\tau_{\sL}^t}},
\qquad
\widetilde{\bR}_{\sL}^t
:=\frac{\bR_{\sL}^t-\bL}{\sqrt{\widehat\tau_{\sS}^t}}.
\end{equation*}
At $t=10$, the entrywise noise quantiles closely match those of
$\Normal(0,1)$ except in the extreme tails
(Fig.~\ref{Fig:Check_Effective_Noise}).
The singular-value distribution of
$\widetilde{\bR}_{\sL}^t/\sqrt n$ also closely matches that of
$\bm{\mathsf G}/\sqrt n$ (Fig.~\ref{Fig:Check_Spectral_Noise}).
Finally, Fig.~\ref{Fig:Check_Observed_Input_Spectrum} compares
the input spectrum of $\bR_{\sL}^t/\sqrt{np_{\sL}}$ with
$\rho_{\widehat\tau_{\sS}^t}$.
In both Figs.~\ref{Fig:Check_Spectral_Noise}
and~\ref{Fig:Check_Observed_Input_Spectrum}, the empirical spectra
closely match the predictions, supporting the Gaussian equivalence
phenomenon in Conjecture~\ref{Conj:Effective}(ii).

\clearpage
\section{Properties of the MSE Transfer Functions}
\label{Appendix:Rectangular_Risk}

In this appendix, we prove Lemmas~\ref{Lemma:Spectral_Ratio}
and~\ref{Lemma:Risk_Conditions}.

\subsection{RIE Shrinkage}
\label{Appendix:Spectral_Supporting_Arguments}

For the Gaussian channel
$\bm{\mathsf R}_\tau=\bL+\sqrt{\tau}\,\bm{\mathsf G}$,
let $\mu_\tau$ be the limiting empirical law of the normalized
singular values
$\sigma_i(\bm{\mathsf R}_\tau)/\sqrt{np_{\sL}}$, $i=1,\ldots,m$.
We write $\rho_\tau$ for its density, $\widehat\mu_\tau$
for its symmetrization, and $\mathcal H_\tau$ for the
Hilbert transform of $\widehat\mu_\tau$. Recall from \eqref{Eq:Optimal_Shrinker} that, for
$s>0$ in $\operatorname{supp}\mu_\tau$,
\begin{equation*}
\hat f_{\beta,\tau}(s)
=s-\frac{\tau}{p_{\sL}}
\left[\frac{1-\beta}{s}+2\beta\mathcal H_\tau(s)\right].
\end{equation*}
Let $\nu$ be the limiting law of the normalized signal singular
values from Appendix~\ref{Appendix:Reference_Conditions}, and set
$M:=\max\operatorname{supp}\nu<\infty$.

\begin{lemma}[Bounds and monotonicity of the RIE shrinker]
\label{Lemma:Noisy_Spectrum_RIE}
For each $\tau>0$, the function $\hat f_{\beta,\tau}$ is
nonnegative, bounded above by $M$, and nondecreasing on
$\operatorname{supp}\mu_\tau$.
\end{lemma}

\begin{proof}[\normalfont\bfseries Proof]
Fix $\tau>0$. We first work at positive singular values with
$\rho_\tau(s)>0$. Recall the Cauchy transform
\begin{equation*}
g_\tau(z):=\int_{\R}\frac{\widehat\mu_\tau(\dd x)}{z-x}.
\end{equation*}
We write $g_\tau(s+\mathrm i0^+)$ for its limit as
$z=s+\mathrm i\eta$ with $\eta\downarrow0$. At such $s$,
the boundary-value formula
\cite[Thm.~2.1]{dozier2007analysis} gives
\begin{equation*}
g_\tau(s+\mathrm i0^+)
=\mathcal H_\tau(s)-\frac{\mathrm i\pi}{2}\rho_\tau(s).
\end{equation*}
Denote
\begin{align*}
q_s&:=s-\frac{\beta\tau}{p_{\sL}}g_\tau(s+\mathrm i0^+),\\
D_s(\xi)&:=q_s\left(q_s-\frac{(1-\beta)\tau}{p_{\sL}s}\right)-\xi^2.
\end{align*}
Then $q_s$ satisfies the following self-consistent equation
\cite[Eq.~(1.2)]{capitaine2014separation}:
\begin{equation}
\frac{p_{\sL}}{\beta\tau}(s-q_s)
=\int\frac{q_s}{D_s(\xi)}\,\nu(\dd\xi).
\label{Eq:RIE_Spectral_Equation}
\end{equation}
Combining the definition of $q_s$ with the shrinkage formula
\eqref{Eq:Optimal_Shrinker} and
$\operatorname{Re}g_\tau(s+\mathrm i0^+)=\mathcal H_\tau(s)$,
we obtain
\begin{equation*}
2\operatorname{Re}q_s
=s+\hat f_{\beta,\tau}(s)+\frac{(1-\beta)\tau}{p_{\sL}s}.
\end{equation*}

\emph{Nonnegativity and boundedness of $\hat f_{\beta,\tau}$.}
We extract a positive weighted average of the squared signal
singular values from \eqref{Eq:RIE_Spectral_Equation}.
Taking imaginary parts, first directly and then after dividing
by $q_s$, gives
\begin{align}
\frac{p_{\sL}}{\beta\tau}
&=\int\frac{|q_s|^2+\xi^2}{|D_s(\xi)|^2}\,\nu(\dd\xi),
\label{Eq:RIE_Spectral_Normalization}\\
\frac{p_{\sL}s}{\beta\tau|q_s|^2}
&=\bigl(s+\hat f_{\beta,\tau}(s)\bigr)
\int\frac{\nu(\dd\xi)}{|D_s(\xi)|^2}.
\nonumber
\end{align}
Combining these identities yields
\begin{equation*}
\frac{\hat f_{\beta,\tau}(s)|q_s|^2}{s}
=\frac{\displaystyle\int\xi^2|D_s(\xi)|^{-2}\,\nu(\dd\xi)}
{\displaystyle\int|D_s(\xi)|^{-2}\,\nu(\dd\xi)}
\in[0,M^2].
\end{equation*}
The ratio is a weighted average of $\xi^2$, with positive
weights and $\xi\in[0,M]$. It follows that
$\hat f_{\beta,\tau}(s)\ge0$. Since $\beta\le1$,
\begin{align*}
|q_s|^2
&\ge\frac14\left(s+\hat f_{\beta,\tau}(s)
+\frac{(1-\beta)\tau}{p_{\sL}s}\right)^2\\
&\ge s\hat f_{\beta,\tau}(s).
\end{align*}
Hence $\hat f_{\beta,\tau}(s)^2\le M^2$, proving the upper bound.

\emph{Monotonicity.}
Fix $0<s<t$ with $\rho_\tau(s)>0$ and $\rho_\tau(t)>0$, and set
\begin{equation*}
\Delta:=q_t-q_s,\qquad
d:=\frac{(1-\beta)\tau}{p_{\sL}}\left(\frac1s-\frac1t\right)\ge0.
\end{equation*}
Subtracting \eqref{Eq:RIE_Spectral_Equation} at $s$ and $t$ gives
\begin{equation*}
\Delta-(t-s)
=\frac{\beta\tau}{p_{\sL}}
\int\frac{q_tq_s(\Delta+d)+\xi^2\Delta}
{D_t(\xi)D_s(\xi)}\,\nu(\dd\xi).
\end{equation*}
Cauchy--Schwarz and \eqref{Eq:RIE_Spectral_Normalization} give
\begin{equation*}
\frac{\beta\tau}{p_{\sL}}
\int\frac{|q_tq_s|+\xi^2}{|D_t(\xi)D_s(\xi)|}\,\nu(\dd\xi)
\le1.
\end{equation*}
The triangle inequality therefore yields
\begin{equation*}
|\Delta-(t-s)|\le\max\{|\Delta+d|,|\Delta|\}.
\end{equation*}
Squaring this inequality gives
$2\operatorname{Re}\Delta\ge t-s$ when the maximum is
$|\Delta|$, and $2\operatorname{Re}\Delta\ge t-s-d$ otherwise.
In either case,
\begin{equation*}
\hat f_{\beta,\tau}(t)-\hat f_{\beta,\tau}(s)
=2\operatorname{Re}\Delta-(t-s)+d\ge0.
\end{equation*}
This comparison also applies when $s$ and $t$ lie in different
components of the spectral support.

For $\tau>0$, the noisy law $\mu_\tau$ has a density on
$(0,\infty)$ and no atom at zero
\cite{dozier2007analysis,capitaine2014separation}.
Thus positive points with $\rho_\tau(s)>0$ are dense in its
support. By the boundary continuity and the value at zero
stated in Appendix~\ref{Appendix:Reference_Conditions}, the bounds
and monotonicity of $\hat f_{\beta,\tau}$ extend to the entire support.
\end{proof}

We use the following bounds to control the spectral integrals
below and the small-noise limit in Appendix~\ref{Appendix:Dimensions}.
The boundary-value formula and Stieltjes-transform estimate in
\cite[Thm.~2.1 and Lemma~2.1(b)]{dozier2007analysis}, after changing
from squared singular values to singular values, give
\begin{equation*}
\|\rho_\tau\|_\infty
\le\frac{2\sqrt{p_{\sL}}}{\pi\sqrt{\beta\tau}}.
\end{equation*}
The singular-value perturbation inequality and the almost-sure
limit $\|\bm{\mathsf G}/\sqrt n\|_{\mathrm{op}}\to1+\sqrt\beta$ give
\begin{equation*}
\operatorname{supp}\mu_\tau
\subset[0,M_\tau],\qquad
M_\tau:=M+(1+\sqrt\beta)\sqrt{\tau/p_{\sL}}.
\end{equation*}

\subsection{MSE Identities}
\label{Appendix:Spectral_MSE_Identities}
For convenience, denote
\begin{equation}
b_\tau(s):=\frac{p_{\sL}}{\tau}
\bigl[s-\hat f_{\beta,\tau}(s)\bigr]
=\frac{1-\beta}{s}+2\beta\mathcal H_\tau(s),
\label{Eq:RIE_Shrinkage_Difference}
\end{equation}
which measures the change in a normalized singular value
under the RIE per unit normalized noise variance $\tau/p_{\sL}$.
The same function determines the motion of the noisy
singular-value distribution as the noise variance changes,
as shown in \eqref{Eq:Rectangular_Spectral_Heat_Flow}.
We denote its mean square by
\begin{align}
J_\beta(\tau)
&:=\int b_\tau(s)^2\,\mu_\tau(\dd s)\nonumber\\
&=\int\left(\frac{1-\beta}{s}\right)^2\mu_\tau(\dd s)
+\frac{\pi^2\beta^2}{3}\int\rho_\tau(s)^3\,\dd s,
\label{Eq:Rectangular_Spectral_Functional}
\end{align}
where the second equality follows from the identities in
\cite[Lemma~5.3]{pourkamali2025rectangular}:
\begin{align*}
\int\mathcal H_\tau(s)^2\,\mu_\tau(\dd s)
&=\frac{\pi^2}{12}\int\rho_\tau(s)^3\,\dd s,\\
\int\frac{\mathcal H_\tau(s)}{s}\,\mu_\tau(\dd s)
&=0,\qquad \beta<1.
\end{align*}

The second moment
$\int s^2\,\mu_\tau(\dd s)=1+\tau/p_{\sL}$ and
\eqref{Eq:Appendix_Spectral_Moments} give
\begin{equation}
\int s b_\tau(s)\,\mu_\tau(\dd s)
=\frac{p_{\sL}}{\tau}
\left(1+\frac{\tau}{p_{\sL}}-1\right)=1.
\label{Eq:Spectral_Shrinkage_Moment}
\end{equation}
Expanding $\hat f_{\beta,\tau}(s)=s-(\tau/p_{\sL})b_\tau(s)$
therefore yields
\begin{equation*}
\int\hat f_{\beta,\tau}(s)^2\,\mu_\tau(\dd s)
=1-\frac{\tau}{p_{\sL}}
+\frac{\tau^2}{p_{\sL}^2}J_\beta(\tau).
\end{equation*}
Using \eqref{Eq:Rectangular_Optimal_Energy} and
\eqref{Eq:Calibration_Optimal_Response}--\eqref{Eq:Calibration_Optimal_Risk},
we obtain the following identities:
\begin{align}
\MMSE_{\sL}(\tau)
&=\tau-\frac{\tau^2}{p_{\sL}}J_\beta(\tau),\nonumber\\
a_{\sL}(\tau)
&=1-\frac{\tau}{p_{\sL}}J_\beta(\tau),
\label{Eq:Rectangular_Risk_Response}\\
\Phi_{\sL}(\tau)
&=\frac{p_{\sL}}{J_\beta(\tau)}-\tau.
\label{Eq:Rectangular_Extrinsic_Functional}
\end{align}

Finally, Cauchy--Schwarz gives
\begin{equation*}
1=\left(\int s b_\tau(s)\,\mu_\tau(\dd s)\right)^2
\le\left(1+\frac{\tau}{p_{\sL}}\right)J_\beta(\tau).
\end{equation*}
Thus $J_\beta(\tau)\ge p_{\sL}/(p_{\sL}+\tau)>0$.
Together with $\MMSE_{\sL}(\tau)\ge0$, this gives
\begin{equation}
0\le a_{\sL}(\tau)
\le\frac{p_{\sL}}{p_{\sL}+\tau}<1.
\label{Eq:Rectangular_Risk_Bounds}
\end{equation}

Lemma~\ref{Lemma:Spectral_Differentiation} below records a useful
differential identity for $J_\beta(\tau)$.

\begin{lemma}
\label{Lemma:Spectral_Differentiation}
The function $J_\beta$ is locally absolutely continuous on
$(0,\infty)$, and for almost every $\tau>0$,
\begin{align}
J_\beta(\tau)+\tau J_\beta'(\tau)
&=\int
\frac{(1-\beta)^2\hat f_{\beta,\tau}(s)}{s^3}\,\mu_\tau(\dd s)\nonumber\\
&\quad+\frac{\pi^2\beta^2}{3}
\int\rho_\tau(s)^3\,\dd\hat f_{\beta,\tau}(s).
\label{Eq:Rectangular_Strict_Ratio_Derivative}
\end{align}
\end{lemma}

\begin{proof}[\normalfont\bfseries Proof]
At each variance $\tau$, the
observation has the same distribution as
$\bm{\mathsf R}_\tau=\bL+\sqrt{\tau}\,\bm{\mathsf G}$.
The large-system limit of the singular-value dynamics
\cite{guionnet2023rectangular} gives the conservation equation
\begin{equation}
\partial_\tau\rho_\tau
=-\frac1{2p_{\sL}}\partial_s(\rho_\tau b_\tau).
\label{Eq:Rectangular_Spectral_Heat_Flow}
\end{equation}
Here $b_\tau(s)/(2p_{\sL})$ is the velocity that transports
the normalized singular-value density as $\tau$ increases.

For $\beta<1$, the support of $\mu_\tau$ is bounded away
from zero for each $\tau>0$
\cite[Prop.~2.3]{capitaine2014separation}.
We may therefore multiply
\eqref{Eq:Rectangular_Spectral_Heat_Flow} by $s^{-2}$ and
integrate by parts to differentiate the inverse-square moment.
For the cubic-density integral, we multiply the same equation
by $3\rho_\tau^2$ and integrate by parts. This gives
\begin{equation}
\begin{aligned}
\frac{\dd}{\dd\tau}\int\frac{\mu_\tau(\dd s)}{s^2}
&=-\frac1{p_{\sL}}
\int\frac{b_\tau(s)}{s^3}\,\mu_\tau(\dd s),\\
\frac{\dd}{\dd\tau}\int\rho_\tau(s)^3\,\dd s
&=-\frac3{2p_{\sL}}\int\rho_\tau(s)^2
\partial_s\bigl(\rho_\tau(s)b_\tau(s)\bigr)\,\dd s\\
&=-\frac1{p_{\sL}}\int\rho_\tau(s)^3
\partial_s b_\tau(s)\,\dd s.
\end{aligned}
\label{Eq:Spectral_Integral_Derivatives}
\end{equation}
Integration by parts produces endpoint contributions involving
$s^{-2}\rho_\tau(s)b_\tau(s)$ for the inverse-square moment
($\beta<1$) and $\rho_\tau(s)^3b_\tau(s)$ for the cubic-density
integral. These expressions vanish at positive spectral endpoints
because $\rho_\tau$ tends to zero, and at infinity because the density
has compact support. At $s=0$, the spectral gap makes $\rho_\tau$
identically zero in a neighborhood of zero when $\beta<1$.
For $\beta=1$, only the cubic-density contribution is present;
it tends to zero because $\rho_\tau$ is bounded and
$b_\tau(s)\to0$ as $s\downarrow0$.
The preceding spectral regularity and bounds justify these
calculations by truncation at the endpoints and give local
absolute continuity of $J_\beta$ for $\tau>0$.

Finally, differentiate $\tau J_\beta(\tau)$ using
\eqref{Eq:Rectangular_Spectral_Functional} and
\eqref{Eq:Spectral_Integral_Derivatives}.
Substituting
$\hat f_{\beta,\tau}(s)=s-(\tau/p_{\sL})b_\tau(s)$ and
$\partial_s\hat f_{\beta,\tau}(s)
=1-(\tau/p_{\sL})\partial_s b_\tau(s)$ gives
\eqref{Eq:Rectangular_Strict_Ratio_Derivative}.
\end{proof}

\subsection{A Differential Inequality for \texorpdfstring{$J_\beta$}{J\_beta}}
Our goal is to show that the spectral MSE transfer
function $\Phi_{\sL}$ is nondecreasing. Since $J_\beta$
is positive and locally absolutely continuous,
$\Phi_{\sL}$ is also locally absolutely continuous.
Differentiating \eqref{Eq:Rectangular_Extrinsic_Functional}
gives
\begin{equation*}
\Phi_{\sL}'(\tau)
=-\frac{p_{\sL}J_\beta'(\tau)}{J_\beta(\tau)^2}-1
\quad\text{for a.e. }\tau>0.
\end{equation*}
The differential inequality in the following lemma
therefore implies $\Phi_{\sL}'(\tau)\ge0$ for almost every
$\tau>0$, establishing the desired monotonicity.

\begin{lemma}[Spectral integral inequality]
\label{Lemma:Spectral_Integral_Inequality}
For almost every $\tau>0$,
\begin{equation*}
J_\beta'(\tau)+\frac{J_\beta(\tau)^2}{p_{\sL}}\le0.
\end{equation*}
\end{lemma}

\begin{proof}[\normalfont\bfseries Proof]
We write \eqref{Eq:RIE_Shrinkage_Difference} as
\begin{equation*}
b_\tau(s)=\frac{1-\beta}{s}
+\beta\,\operatorname{p.v.}\!\int
\left(\frac1{s-t}+\frac1{s+t}\right)\mu_\tau(\dd t).
\end{equation*}
For a fixed probability level $u\in(0,1)$, define
\begin{equation*}
\begin{aligned}
F_\tau(s)&:=\int_0^s\rho_\tau(x)\,\dd x,\\
s_u(\tau)&:=\inf\{s\ge0:F_\tau(s)\ge u\}.
\end{aligned}
\end{equation*}
Thus the probability mass below $s_u(\tau)$ is $u$.
As $\tau$ varies, we hold $u$ fixed and follow the corresponding
singular-value location $s_u(\tau)$. Differentiating
$F_\tau(s_u(\tau))=u$ and using
\eqref{Eq:Rectangular_Spectral_Heat_Flow}, together with the
vanishing boundary contribution at zero, gives
\begin{equation*}
\frac{\dd s_u(\tau)}{\dd\tau}
=\frac{b_\tau(s_u(\tau))}{2p_{\sL}}
\end{equation*}
at points where $\rho_\tau(s_u(\tau))>0$.
For two fixed probability levels $u,v\in(0,1)$, write
$s=s_u(\tau)$ and $t=s_v(\tau)$. Then
\begin{equation*}
\frac{\dd}{\dd\tau}\frac1{s\pm t}
=-\frac{b_\tau(s)\pm b_\tau(t)}
{2p_{\sL}(s\pm t)^2}.
\end{equation*}
This parametrization also gives
\begin{equation*}
J_\beta(\tau)
=\int_0^1 b_\tau(s_u(\tau))^2\,\dd u.
\end{equation*}
We differentiate with respect to $\tau$ while keeping
the probability levels fixed. We then average each resulting
double integral with the expression obtained by exchanging
$u$ and $v$. Expressing the result in terms of $s$ and $t$
gives
\begin{align*}
&-p_{\sL}J_\beta'(\tau)\\
&\quad=\int(1-\beta)
\left(\frac{b_\tau(s)}s\right)^2\mu_\tau(\dd s)\\
&\qquad+\frac\beta2\iint
\left(\frac{b_\tau(s)-b_\tau(t)}{s-t}\right)^2
\mu_\tau(\dd s)\mu_\tau(\dd t)\\
&\qquad+\frac\beta2\iint
\left(\frac{b_\tau(s)+b_\tau(t)}{s+t}\right)^2
\mu_\tau(\dd s)\mu_\tau(\dd t).
\end{align*}
On the other hand, multiplying the expression for $b_\tau$
by $b_\tau(s)$ and symmetrizing gives the same weighted
integrals without the squares:
\begin{align*}
J_\beta(\tau)
&=\int(1-\beta)\frac{b_\tau(s)}s\,\mu_\tau(\dd s)\\
&\quad+\frac\beta2\iint
\frac{b_\tau(s)-b_\tau(t)}{s-t}\,
\mu_\tau(\dd s)\mu_\tau(\dd t)\\
&\quad+\frac\beta2\iint
\frac{b_\tau(s)+b_\tau(t)}{s+t}\,
\mu_\tau(\dd s)\mu_\tau(\dd t).
\end{align*}
The weights $1-\beta,\beta/2,\beta/2$ sum to one, and
$\mu_\tau$ is a probability measure. Cauchy--Schwarz thus gives
$J_\beta(\tau)^2\le-p_{\sL}J_\beta'(\tau)$.
The singular kernel $1/(s-t)$ in the expression for $b_\tau$
is interpreted in the principal-value sense.
For the difference quotients in the symmetrized integrals,
the value at $t=s$ is obtained from
\begin{equation*}
\lim_{t\to s}
\frac{b_\tau(s)-b_\tau(t)}{s-t}
=\partial_s b_\tau(s),
\end{equation*}
wherever this derivative exists.
The preceding spectral regularity justifies the
differentiations and limiting operations, with the resulting
derivative identities holding for almost every $\tau>0$.
\end{proof}

\subsection{Proofs of the MSE Properties}
\label{Appendix:Extrinsic_Monotonicity}

\begin{proof}[\normalfont\bfseries Proof of Lemma~\ref{Lemma:Spectral_Ratio}]
Equations~\eqref{Eq:Rectangular_Risk_Response} and
\eqref{Eq:Rectangular_Strict_Ratio_Derivative} give
\begin{equation*}
-a_{\sL}'(\tau)
=\frac{J_\beta(\tau)+\tau J_\beta'(\tau)}{p_{\sL}}\ge0
\quad\text{for a.e. }\tau>0,
\end{equation*}
because $\hat f_{\beta,\tau}\ge0$ and
$\partial_s\hat f_{\beta,\tau}\ge0$ where $\rho_\tau>0$.
To obtain strict decrease, we show that at least one term
in \eqref{Eq:Rectangular_Strict_Ratio_Derivative} is positive.

For $\beta<1$, the first term is positive because
$\hat f_{\beta,\tau}\ge0$ and
$\int s\hat f_{\beta,\tau}(s)\,\mu_\tau(\dd s)=1$
by \eqref{Eq:Appendix_Spectral_Moments}.
For $\beta=1$, only the second term remains, so we show that
the shrinker is strictly increasing where $\rho_\tau>0$.
Recall
$q_s=s-(\tau/p_{\sL})g_\tau(s+\mathrm i0^+)$
from the proof of Lemma~\ref{Lemma:Noisy_Spectrum_RIE}.
Differentiating \eqref{Eq:RIE_Spectral_Equation} gives
\begin{equation*}
q_s'-1
=\frac{\tau}{p_{\sL}}q_s'
\int\frac{q_s^2+\xi^2}{(q_s^2-\xi^2)^2}\,\nu(\dd\xi).
\end{equation*}
Since $\operatorname{Im}q_s>0$, we have
$|q_s^2+\xi^2|<|q_s|^2+\xi^2$ for $\xi>0$.
For fixed $\tau$, define
$q(z):=z-(\tau/p_{\sL})g_\tau(z)$ for
$\operatorname{Im}z>0$. Thus $q_s=q(s+\mathrm i0^+)$.
The prime in $q'(z)$ denotes differentiation with respect to $z$,
and $s+\mathrm i0^+$ denotes the limit $z=s+\mathrm i\eta$
as $\eta\downarrow0$.
Where $\rho_\tau(s)>0$, boundary regularity gives
$q'(s+\mathrm i0^+)=\partial_s q(s+\mathrm i0^+)$
\cite{biane1997semicircular}.

The unit second moment of $\nu$ and
\eqref{Eq:RIE_Spectral_Normalization} therefore give
\begin{align*}
&\bigl|q'(s+\mathrm i0^+)-1\bigr|\\
&\quad<\frac{\tau}{p_{\sL}}
\bigl|q'(s+\mathrm i0^+)\bigr|\\
&\qquad{}\times
\int\frac{|q(s+\mathrm i0^+)|^2+\xi^2}
{|q(s+\mathrm i0^+)^2-\xi^2|^2}\,\nu(\dd\xi)\\
&\quad=\bigl|q'(s+\mathrm i0^+)\bigr|.
\end{align*}
Squaring yields $2\operatorname{Re}q'(s+\mathrm i0^+)>1$.
Since $\hat f_{1,\tau}(s)=2\operatorname{Re}q(s+\mathrm i0^+)-s$,
we obtain
\begin{equation*}
\partial_s\hat f_{1,\tau}(s)
=2\operatorname{Re}q'(s+\mathrm i0^+)-1>0.
\end{equation*}
This makes the remaining term in
\eqref{Eq:Rectangular_Strict_Ratio_Derivative} positive.

Integrating $-a_{\sL}'(\tau)>0$ shows that $a_{\sL}$ is
strictly decreasing. Together with
\eqref{Eq:Rectangular_Risk_Bounds}, this gives
$0<a_{\sL}(\tau)\le p_{\sL}/(p_{\sL}+\tau)$ and
$a_{\sL}(\tau)\to0$ as $\tau\to\infty$.
Its limit $d_{\sL}=a_{\sL}(0+)$ thus exists in $(0,1]$.
Setting $a_{\sL}(0)=d_{\sL}$ gives the claimed continuous,
strictly decreasing extension.
\end{proof}

\begin{proof}[\normalfont\bfseries Proof of Lemma~\ref{Lemma:Risk_Conditions}]
For the spectral module,
\eqref{Eq:Rectangular_Extrinsic_Functional} and
Lemma~\ref{Lemma:Spectral_Integral_Inequality} give
\begin{equation*}
\Phi_{\sL}'(\tau)
=-\frac{p_{\sL}J_\beta'(\tau)}{J_\beta(\tau)^2}-1\ge0
\quad\text{for a.e. }\tau>0.
\end{equation*}
Integrating this inequality shows that $\Phi_{\sL}$
is nondecreasing.

For the entrywise channel
$\mathsf R_\tau=\mathsf S+\sqrt{\tau}\,\mathsf Z$,
let $V_\tau(r)=\operatorname{Var}(\mathsf S\mid\mathsf R_\tau=r)$
be the posterior variance.
The Gaussian MMSE identity
\cite[Props.~7 and~9]{guo2011mmseproperties} gives
\begin{equation*}
\MMSE_{\sS}(\tau)=\E[V_\tau(\mathsf R_\tau)],
\qquad
\MMSE_{\sS}'(\tau)
=\frac{\E[V_\tau(\mathsf R_\tau)^2]}{\tau^2}.
\end{equation*}
Differentiating
$\Phi_{\sS}(\tau)=\tau\MMSE_{\sS}(\tau)/
(\tau-\MMSE_{\sS}(\tau))$ then yields
\begin{align*}
\Phi_{\sS}'(\tau)
&=\frac{\tau^2\MMSE_{\sS}'(\tau)-\MMSE_{\sS}(\tau)^2}
{(\tau-\MMSE_{\sS}(\tau))^2}\\
&=\frac{\operatorname{Var}(V_\tau(\mathsf R_\tau))}
{(\tau-\MMSE_{\sS}(\tau))^2}\ge0.
\end{align*}
Thus the variation of the posterior variance across
observations determines how fast $\Phi_{\sS}$ increases.

For both modules, the MMSE is positive and continuous
for $\tau>0$, and the linear-estimation bound gives
\begin{equation*}
0<\MMSE_\bullet(\tau)
\le\frac{p_\bullet\tau}{p_\bullet+\tau}<\tau,
\qquad \bullet\in\{\sL,\sS\}.
\end{equation*}
Recall the normalized functions
\begin{align*}
\bar a_\bullet(x)&=a_\bullet(p_\bullet x),\\
\bar\Psi_\bullet(x)
&=\frac{\Phi_\bullet(p_\bullet x)}{p_\bullet}
=\frac{x\bar a_\bullet(x)}{1-\bar a_\bullet(x)}.
\end{align*}
These formulas preserve continuity and the monotonicity
of the transfer functions. They also give
$0<\bar a_\bullet(x)\le(1+x)^{-1}$,
$0<\bar\Psi_\bullet(x)\le1$, and
$\bar a_\bullet(x)\to0$ as $x\to\infty$.

Finally, if $\bar a_\bullet(x)\to d_\bullet<1$
as $x\downarrow0$, then
\begin{equation}
\bar\Psi_\bullet(x)
=\frac{x\bar a_\bullet(x)}{1-\bar a_\bullet(x)}
=\frac{d_\bullet}{1-d_\bullet}x+o(x).
\label{Eq:Reference_Extrinsic_Small_Noise_Proof}
\end{equation}
Hence $\bar\Psi_\bullet$ extends continuously to
$\bar\Psi_\bullet(0)=0$.
\end{proof}

\section{Derivations of MMSE Dimensions}
\label{Appendix:Dimensions}

We compute $d_{\sS}$ and $d_{\sL}$ in
Theorem~\ref{Theorem:Power_Regimes}. The former is the scalar
MMSE dimension \cite{Wu2011MMSE}; we call the latter the
\emph{spectral MMSE dimension}.

\subsection{Scalar MMSE Dimension}

For the unit-power prior in \eqref{Eqn:Ps_mix}, finite second
moment and \cite[Thm.~15]{Wu2011MMSE} give the following result.
Here, $\sigma^2$ denotes the normalized noise variance.

\begin{lemma}[Scalar MMSE dimension]
For the prior in \eqref{Eqn:Ps_mix},
\begin{equation*}
d_{\sS}=\lim_{\sigma\downarrow0}
\frac{\bar\MMSE_{\sS}(\sigma^2)}{\sigma^2}
=\rho_{\sS}.
\end{equation*}
\end{lemma}

\subsection{Spectral MMSE Dimension}
For the spectral channel, write
$\bL^\circ:=\widetilde{\bL}/\sqrt n=\bL/\sqrt{np_{\sL}}$;
its limiting singular-value law is $\nu$,
as defined in Appendix~\ref{Appendix:Reference_Conditions}.
Since $\bar\MMSE_{\sL}$ is scale-invariant, we set
$p_{\sL}=1$ without loss of generality.
Throughout this subsection, $\sigma$ denotes the noise
standard deviation and $\sigma^2$ its variance.
Thus $\tau=\sigma^2$ in the notation of Appendix~\ref{Appendix:Rectangular_Risk},
and $a_{\sL}(\sigma^2)=\bar a_{\sL}(\sigma^2)$.
For $\sigma>0$, set
\begin{equation*}
\bY_\sigma=\bL^\circ+\sigma\widetilde{\bG},
\qquad \widetilde G_{ij}\overset{\mathrm{i.i.d.}}{\sim}\Normal(0,1/n),
\end{equation*}
where $\widetilde{\bG}$ is independent of the signal.
The limiting singular-value law of $\bY_\sigma$ is therefore
$\mu_{\sigma^2}$, with symmetrization $\widehat\mu_{\sigma^2}$.
We use the atom masses $w_0,w_\xi$ and the set $\mathcal A$ from
Section~\ref{Sec:Simultaneous_Recovery}, and define the nonzero
spectral mass $\gamma_\nu:=1-w_0$.

\begin{theorem}[Spectral MMSE dimension]
\label{Theorem:Spectral_MMSE_Dimension}
The limit
\begin{equation*}
d_{\sL}=\lim_{\sigma\downarrow0}
\frac{\bar\MMSE_{\sL}(\sigma^2)}{\sigma^2}
\end{equation*}
exists and satisfies
\begin{align}
d_{\sL}
&=1-(1-\beta)w_0-\beta w_0^2
-\frac\beta2\sum_{\xi\in\mathcal A}w_\xi^2\nonumber\\
&=\gamma_\nu(1+\beta-\beta\gamma_\nu)
-\frac\beta2\sum_{\xi\in\mathcal A}w_\xi^2.
\label{Eq:GP_Gram_Atom_Formula}
\end{align}
\end{theorem}

We determine $d_{\sL}$ by relating the spectral MMSE to the
logarithmic energy of the singular-value law of the signal-plus-noise matrix and
computing the leading small-noise term of this energy from
the atoms of $\nu$. We establish the energy-derivative identity
and Lemma~\ref{Lemma:Log_Energy_MMSE} below before proving the theorem.

Define the logarithmic energy using the symmetrized
singular-value law \cite[Result~2]{troiani2022optimal}:
\begin{align}
{E_\beta(\sigma^2)}
&{ :=(1-\beta)\int\log s\,\mu_{\sigma^2}(\dd s)}\nonumber\\
&{\quad+\beta\iint\log|x-y|\,
\widehat\mu_{\sigma^2}(\dd x)\widehat\mu_{\sigma^2}(\dd y).}
\label{Eq:Rectangular_Log_Energy}
\end{align}
Averaging over the signs in $\widehat\mu_{\sigma^2}$
converts the interaction kernel $\log|x-y|$ to
$\tfrac12\log|s^2-t^2|$ for nonnegative singular values $s,t$.
For each $\sigma>0$, both $\mu_{\sigma^2}$ and
$\widehat\mu_{\sigma^2}$ have bounded densities and compact
support by Appendix~\ref{Appendix:Spectral_Supporting_Arguments}.
Since the logarithmic singularities are locally integrable,
both integrals in \eqref{Eq:Rectangular_Log_Energy} are finite.

As in the proof of
Lemma~\ref{Lemma:Spectral_Integral_Inequality}, follow
singular-value locations with fixed cumulative probabilities.
With respect to $\sigma$, these locations move with velocity
$\sigma b_{\sigma^2}$.
The same regularity arguments show that
$\sigma\mapsto E_\beta(\sigma^2)$ is locally absolutely
continuous on $(0,\infty)$.
Differentiating the energy along this motion and using
\eqref{Eq:Rectangular_Risk_Response} gives, for almost every
$\sigma>0$,
\begin{align}
{\frac{\dd}{\dd\sigma}E_\beta(\sigma^2)}
&{=\sigma\int b_{\sigma^2}(s)^2\,
\mu_{\sigma^2}(\dd s)}\nonumber\\
&{=\sigma J_\beta(\sigma^2)
=\frac{1-\bar a_{\sL}(\sigma^2)}{\sigma}.}
\label{Eq:Log_Energy_Derivative}
\end{align}

By Lemma~\ref{Lemma:Spectral_Ratio}, $\bar a_{\sL}(0+)=d_{\sL}$.
The following lemma relates this limit to the logarithmic energy.

\begin{lemma}[Logarithmic-energy identity]
\label{Lemma:Log_Energy_MMSE}
For $0<\sigma<\sigma_0$,
\begin{equation}
E_\beta(\sigma_0^2)-E_\beta(\sigma^2)
=\int_\sigma^{\sigma_0}
\frac{1-\bar a_{\sL}(t^2)}{t}\,\dd t.
\label{Eq:Log_Energy_Increment}
\end{equation}
Moreover,
\begin{align}
\lim_{\sigma\downarrow0}
\frac{E_\beta(\sigma^2)}{\log\sigma}
&=1-d_{\sL}
\label{Eq:Energy_MMSE_Dimension}\\
&=(1-\beta)w_0+\beta w_0^2
+\frac{\beta}{2}\sum_{\xi\in\mathcal A}w_\xi^2.
\label{Eq:Compact_Atom_Energy_Slope}
\end{align}
\end{lemma}

\begin{proof}[\normalfont\bfseries Proof]
Integrating \eqref{Eq:Log_Energy_Derivative}
from $\sigma$ to $\sigma_0$ gives
\eqref{Eq:Log_Energy_Increment}.
The limit $\bar a_{\sL}(0+)=d_{\sL}$ then yields
\eqref{Eq:Energy_MMSE_Dimension}.
It remains to prove \eqref{Eq:Compact_Atom_Energy_Slope}.

To regularize the logarithmic singularity at zero at the noise scale $\sigma\in(0,1)$, define
\begin{equation*}
\ell_\sigma(z)=\min\{\log\max\{|z|,\sigma\},0\},
\qquad
\chi_\sigma(z)=\frac{\ell_\sigma(z)}{\log\sigma}.
\end{equation*}
For any probability measure $\mu$ on $\R_+$ with symmetrization $\widehat\mu$, define the truncated functional
\begin{equation*}
\begin{aligned}
\mathcal{Q}_\sigma(\mu)
&=(1-\beta)\int \chi_\sigma(s)\,\mu(\dd s)\\
&\quad+\beta\iint \chi_\sigma(x-y)\,\widehat\mu(\dd x)\widehat\mu(\dd y).
\end{aligned}
\end{equation*}
As $\sigma\downarrow0$, the claimed limit follows from the three steps
\begin{align*}
\frac{E_\beta(\sigma^2)}{\log\sigma}
&\overset{\text{(a)}}{=}\mathcal{Q}_\sigma(\mu_{\sigma^2})+o(1)
\overset{\text{(b)}}{=}\mathcal{Q}_\sigma(\nu)+o(1)\\
&\overset{\text{(c)}}{\longrightarrow}
(1-\beta)w_0+\beta w_0^2
+\frac{\beta}{2}\sum_{\xi\in\mathcal A}w_\xi^2.
\end{align*}

\emph{Step (a).} By Appendix~\ref{Appendix:Spectral_Supporting_Arguments}, there exist constants $M_*>1$ and $C_\rho>0$, independent of $\sigma\in(0,1)$, such that $\mu_{\sigma^2}$ is supported on $[0,M_*]$ with density bounded by $C_\rho/\sigma$. Under $\widehat\mu_{\sigma^2}\otimes\widehat\mu_{\sigma^2}$, the difference $x-y$ is supported on $[-2M_*,2M_*]$ with convolution density also bounded by $C_\rho/\sigma$. On $[-2M_*,2M_*]\setminus\{0\}$,
\begin{equation*}
|\log|z|-\ell_\sigma(z)|
\le \mathbf{1}_{\{|z|<\sigma\}}\log(\sigma/|z|)+\log(2M_*).
\end{equation*}
Integrating against a density bounded by $C_\rho/\sigma$ bounds the truncation error in either integral of $E_\beta(\sigma^2)$ by
\begin{equation*}
\frac{2C_\rho}{\sigma}\int_0^\sigma\log(\sigma/t)\,\dd t+\log(2M_*)
=2C_\rho+\log(2M_*).
\end{equation*}
Since this bound is uniform in $\sigma\in(0,1)$, dividing by $\log\sigma$ and letting $\sigma\downarrow0$ yields (a).

\emph{Step (b).} Recall that $\bY_\sigma=\bL^\circ+\sigma\widetilde{\bG}$, where $\widetilde{\bG}$ has i.i.d.\ $\Normal(0,1/n)$ entries independently of $\bL^\circ$. Write $s_{\sigma,i}=s_i(\bY_\sigma)$ and $s_{0,i}=s_i(\bL^\circ)$ for their singular values in nonincreasing order, and let $\mu_{\sigma^2,n}$ and $\nu_n$ denote their empirical distributions. By Weyl's inequality, $|s_{\sigma,i}-s_{0,i}|\le\sigma\|\widetilde{\bG}\|_{\mathrm{op}}$ for all $i\in[m]$. Since $\ell_\sigma$ is $1/\sigma$-Lipschitz,
\begin{equation*}
\left|\frac1m\sum_{i=1}^m
\bigl[\ell_\sigma(s_{\sigma,i})-\ell_\sigma(s_{0,i})\bigr]\right|
\le\|\widetilde{\bG}\|_{\mathrm{op}}.
\end{equation*}
Similarly, for every $i,j\in[m]$ and $\varsigma_1,\varsigma_2\in\{-1,1\}$,
\begin{equation*}
\begin{aligned}
&\bigl|\ell_\sigma(\varsigma_1 s_{\sigma,i}-\varsigma_2 s_{\sigma,j})
-\ell_\sigma(\varsigma_1 s_{0,i}-\varsigma_2 s_{0,j})\bigr|\\
&\qquad\le\frac{|s_{\sigma,i}-s_{0,i}|+|s_{\sigma,j}-s_{0,j}|}{\sigma}
\le2\|\widetilde{\bG}\|_{\mathrm{op}}.
\end{aligned}
\end{equation*}
Averaging over $i,j\in[m]$ and $\varsigma_1,\varsigma_2\in\{-1,1\}$ bounds the difference between the double integrals of $\ell_\sigma(x-y)$ under $\widehat\mu_{\sigma^2,n}$ and $\widehat\nu_n$ by $2\|\widetilde{\bG}\|_{\mathrm{op}}$. For each fixed $\sigma\in(0,1)$, $\ell_\sigma$ is bounded and continuous, so letting $m,n\to\infty$ passes these empirical averages to the limiting integrals under $\mu_{\sigma^2},\nu$ and $\widehat\mu_{\sigma^2},\widehat\nu$ while $\|\widetilde{\bG}\|_{\mathrm{op}}\to1+\sqrt\beta$ almost surely. Combining the single and double integrals with $\chi_\sigma=\ell_\sigma/\log\sigma$ gives
\begin{align*}
\bigl|\mathcal{Q}_\sigma(\mu_{\sigma^2})-\mathcal{Q}_\sigma(\nu)\bigr|
&\le\frac{(1-\beta)(1+\sqrt\beta)+2\beta(1+\sqrt\beta)}{|\log\sigma|}\\
&=\frac{(1+\beta)(1+\sqrt\beta)}{|\log\sigma|},
\end{align*}
which tends to zero as $\sigma\downarrow0$ and establishes (b).

\emph{Step (c).} As $\sigma\downarrow0$, $\chi_\sigma(z)\in[0,1]$ converges pointwise to $\mathbf{1}_{\{z=0\}}$. Bounded convergence gives \[\int \chi_\sigma(s)\,\nu(\dd s)\to\nu(\{0\})=w_0\] and passes the double integral to the diagonal mass of $\widehat\nu\otimes\widehat\nu$. Since $\widehat\nu$ assigns mass $w_0$ to $0$ and mass $w_\xi/2$ to each of $\pm\sqrt{\xi}$ for $\xi\in\mathcal A$,
\begin{equation}
\iint\mathbf{1}_{\{x=y\}}\,\widehat\nu(\dd x)\widehat\nu(\dd y)
=w_0^2+\frac12\sum_{\xi\in\mathcal A}w_\xi^2.
\end{equation}
Substituting these limits into $\mathcal{Q}_\sigma(\nu)$ gives (c).
\end{proof}

\begin{proof}[\normalfont\bfseries Proof of Theorem~\ref{Theorem:Spectral_MMSE_Dimension}]
The existence of $d_{\sL}$ follows from
Lemma~\ref{Lemma:Spectral_Ratio}. Equating the limits in
Lemma~\ref{Lemma:Log_Energy_MMSE}
gives
\begin{equation*}
1-d_{\sL}=(1-\beta)w_0+\beta w_0^2
+\frac\beta2\sum_{\xi\in\mathcal A}w_\xi^2.
\end{equation*}
Using $w_0=1-\gamma_\nu$ yields
\eqref{Eq:GP_Gram_Atom_Formula}.
\end{proof}

\begin{figure}[!t]
\centering
\setlength{\abovecaptionskip}{4pt}
\includegraphics[width=\columnwidth]{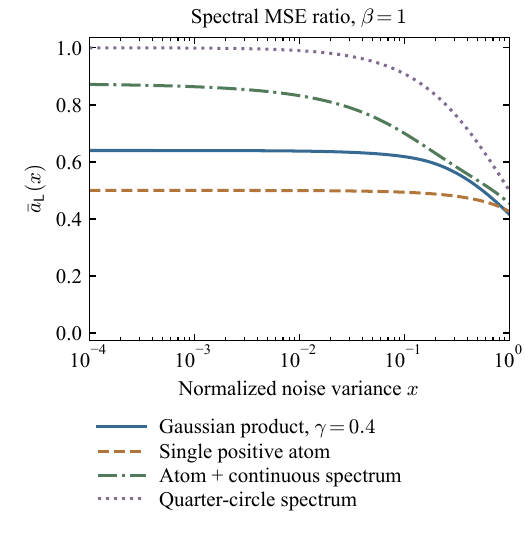}\par\nointerlineskip
\caption{Denoising errors normalized by the noise variance for
unit-power spectral priors at $\beta=1$, plotted against $x=\sigma^2$.
The examples are
the Gaussian-product model with $\gamma=0.4$;
the single-atom law $\delta_1$;
the mixture
$\tfrac12\delta_{\sqrt{11/12}}
+\tfrac12\operatorname{Unif}[1/2,3/2]$;
and the quarter-circle law with density
$\pi^{-1}\sqrt{4-s^2}$ on $[0,2]$.
Their spectral MMSE dimensions are
$0.64$, $0.5$, $0.875$, and $1$, respectively.}
\label{Fig:Reference_Ratios}
\end{figure}

\begin{figure}[!t]
\centering
\setlength{\abovecaptionskip}{4pt}
\includegraphics[width=\columnwidth]{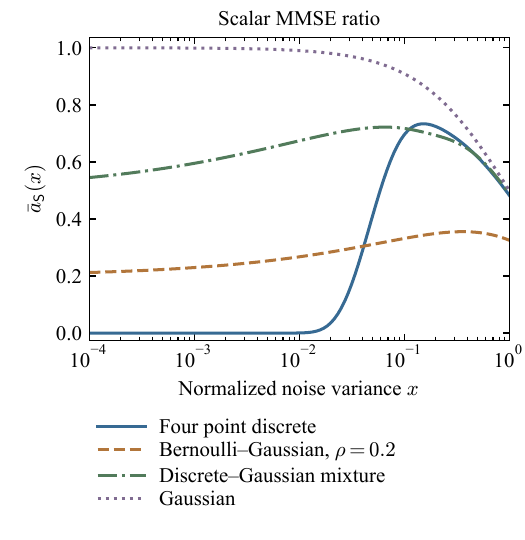}\par\nointerlineskip
\caption{Denoising errors normalized by the noise variance for
unit-power scalar priors, plotted against $x=\sigma^2$.
The examples are
the uniform four-point discrete prior on
$\{-3,-1,1,3\}/\sqrt5$;
the Bernoulli--Gaussian prior
$0.8\delta_0+0.2\Normal(0,5)$;
the mixture
$\tfrac14\delta_{-1}+\tfrac14\delta_1
+\tfrac12\Normal(0,1)$;
and the standard Gaussian prior $\Normal(0,1)$.
Their scalar MMSE dimensions are
$0$, $0.2$, $0.5$, and $1$, respectively.}
\label{Fig:Reference_Scalar_Ratios}
\end{figure}

\subsection{Examples of the Spectral MMSE Dimension}
\label{Appendix:Prior_Examples}

We compare the spectral MMSE dimensions of two matrix
priors with the same limiting rank fraction.

\emph{Gaussian-product prior.}
Let $\bL=\sqrt{p_{\sL}/r}\,\bA\bB$, where
$\bA\in\R^{m\times r}$ and
$\bB\in\R^{r\times n}$ are independent matrices
with i.i.d.\ $\Normal(0,1)$ entries.
Assume $m/n\to\beta\in(0,1]$ and
$r/m\to\gamma\in(0,1)$.
The limiting singular-value law of
$\bL^\circ$
has mass $1-\gamma$ at zero and no positive atoms
\cite{dupic2014products}.
Theorem~\ref{Theorem:Spectral_MMSE_Dimension} gives
\begin{equation}
d_{\sL}
=\gamma(1+\beta-\beta\gamma)
=\lim_{m,n\to\infty}
\frac{r(m+n-r)}{mn}.
\label{Eq:GP_Dimension}
\end{equation}
Since the manifold of real rank-$r$ matrices has
dimension $r(m+n-r)$
\cite[Prop.~2.1]{vandereycken2013lowrank},
$d_{\sL}$ equals its asymptotic dimension per matrix entry.
Under the theorem's assumptions, any limiting law
with mass $1-\gamma$ at zero and no positive atoms
gives the same $d_{\sL}$.

\emph{Equal nonzero singular values.}
For comparison, consider a rotationally invariant prior
for $\bL^\circ$ whose $r$ nonzero singular values
are all fixed at $\sqrt{m/r}$.
With the same limits for $m/n$ and $r/m$, its limiting
singular-value law is
$(1-\gamma)\delta_0+\gamma\delta_{1/\sqrt\gamma}$.
Theorem~\ref{Theorem:Spectral_MMSE_Dimension} now gives
\begin{align*}
d_{\sL}
&=\gamma(1+\beta-\beta\gamma)
-\frac{\beta\gamma^2}{2}\\
&=\lim_{m,n\to\infty}
\frac{r(m+n-r)-r(r+1)/2}{mn}.
\end{align*}
The numerator is the dimension of the manifold of
real $m\times n$ matrices with these fixed singular
values \cite[Ch.~3, Prop.~2.1]{helmke1996optimization}.
Thus, relative to \eqref{Eq:GP_Dimension}, the decrease
$\beta\gamma^2/2$ in $d_{\sL}$ agrees with the decrease
in asymptotic dimension per entry.
Fixing the $r$ nonzero singular values removes $r$ degrees of freedom.
Since these values are equal, applying the same orthogonal rotation
to the $r$ left and right singular vectors leaves $\bL^\circ$ unchanged,
removing another $r(r-1)/2$ degrees of freedom.
The total reduction is therefore $r+r(r-1)/2=r(r+1)/2$.

\subsection{\texorpdfstring{Examples of $\bar a_{\sL}(\sigma^2)$ and $\bar a_{\sS}(\sigma^2)$}{Examples of the spectral and scalar MSE ratios}}

Figures~\ref{Fig:Reference_Ratios} and~\ref{Fig:Reference_Scalar_Ratios} plot
$\bar a_{\sL}(\sigma^2)$ and $\bar a_{\sS}(\sigma^2)$
for several unit-power priors, respectively.
Their limits as $\sigma^2\downarrow0$ are
$d_{\sL}$ and $d_{\sS}$, respectively.
In Fig.~\ref{Fig:Reference_Ratios}, $\bar a_{\sL}(\sigma^2)$ decreases
strictly with $\sigma^2$, as established in
Lemma~\ref{Lemma:Spectral_Ratio}.
In contrast, for each of the three discrete or mixed
scalar priors in Fig.~\ref{Fig:Reference_Scalar_Ratios},
$\bar a_{\sS}(\sigma^2)$ is nonmonotone and exceeds
its zero-noise limit $d_{\sS}$ at some $\sigma^2>0$.
Hence, for each of these priors,
\[
\Delta_{\sS}
=\sup_{x>0}\bar a_{\sS}(x)-d_{\sS}>0.
\]

\section{Convergence and Recovery of State Evolution}

We show that the SE recursion \eqref{Eq:SE_Recursion},
initialized at $\tau_{\sL}^1=p_{\sL}$, converges to its
largest fixed point and derive a necessary and sufficient
condition for perfect recovery.
We use this condition to prove
Theorem~\ref{Theorem:Power_Regimes}, then specialize
the results to Gaussian-product signals and
Bernoulli--Gaussian noise.

Throughout, we fix the aspect ratio $\beta\in(0,1]$
and the unit-power component laws under the assumptions
of Lemmas~\ref{Lemma:Spectral_Ratio}
and~\ref{Lemma:Risk_Conditions}, with
$0<d_{\sL}<1$ and $0\le d_{\sS}<1$.

\subsection{Convergence of the SE Recursion}
We establish convergence from the initialization
$\tau_{\sL}^1=p_{\sL}$ and determine when both SE
error variances converge to zero.
With $\zeta$ defined in Section~\ref{Sec:Simultaneous_Recovery}, let
$x^t=\tau_{\sL}^t/p_{\sS}$, so that $x^1=\zeta^{-1}$.
Combining the two updates in \eqref{Eq:SE_Recursion} gives
\begin{equation}
x^{t+1}
=\frac{1}{\zeta}\,
\bar\Psi_{\sL}\!\left(\zeta\bar\Psi_{\sS}(x^t)\right)
=:\bar\Psi(x^t).
\end{equation}

By Lemma~\ref{Lemma:Risk_Conditions},
$\bar\Psi$ is continuous and nondecreasing, with
$\bar\Psi(0)=0$ and
$0<\bar\Psi(x)\le\zeta^{-1}$ for $x>0$.
In particular,
$x^2=\bar\Psi(x^1)\le\zeta^{-1}=x^1$.
Monotonicity then gives
$0\le x^{t+1}\le x^t$ for every $t\ge1$.
Thus $x^t$ converges to a limit $x_*$, and continuity
gives $\bar\Psi(x_*)=x_*$.
Every fixed point is at most $x^1$ and, by monotonicity,
remains below every iterate.
Hence $x_*$ is the largest fixed point of $\bar\Psi$.

The second error variance satisfies
$\tau_{\sS}^t=p_{\sS}\bar\Psi_{\sS}(x^t)$.
Since $\bar\Psi_{\sS}$ is continuous with
$\bar\Psi_{\sS}(0)=0$, both SE error variances converge
to zero if and only if $x_*=0$.
This is equivalent to
$\bar\Psi(x)<x$ for every $x>0$.
Indeed, if this inequality fails, continuity and
$\bar\Psi(u)<u$ for $u>\zeta^{-1}$ imply a positive
fixed point.
Using
$\bar\Psi_\bullet(u)
=u\bar a_\bullet(u)/(1-\bar a_\bullet(u))$, we obtain
\[
\frac{\bar\Psi(x)}x
=\frac{\bar a_{\sS}(x)}{1-\bar a_{\sS}(x)}
\frac{\bar a_{\sL}(\zeta\bar\Psi_{\sS}(x))}
{1-\bar a_{\sL}(\zeta\bar\Psi_{\sS}(x))}.
\]
Since both denominators are positive, $\bar\Psi(x)<x$ is equivalent to
\begin{equation}
\bar a_{\sS}(x)
+\bar a_{\sL}\!\left(\zeta\bar\Psi_{\sS}(x)\right)<1
\label{Eq:Global_Recovery_Criterion}
\end{equation}
for every $x>0$.

If the spectral update is performed first, take
$\tau_{\sS}^0=p_{\sS}$.
The sequence $\{\tau_{\sS}^t/p_{\sS}\}$ starts at $1$
and is generated by a continuous, nondecreasing map
bounded above by $1$.
It therefore converges to the largest fixed point
of that map.
Since the two update orders have the same fixed-point
equations, both satisfy the recovery criterion
\eqref{Eq:Global_Recovery_Criterion}.

\subsection{Proof of Theorem~\ref{Theorem:Power_Regimes}}
\label{Appendix:Simultaneous_Recovery}
\begin{proof}[\normalfont\bfseries Proof]
For each $x>0$, define
\begin{equation*}
\zeta_{\mathrm{req}}(x):=
\frac{\bar a_{\sL}^{-1}(1-\bar a_{\sS}(x))}
     {\bar\Psi_{\sS}(x)},
\end{equation*}
using the extended inverse in the theorem.
Strict decrease of $\bar a_{\sL}$ and positivity of
$\bar\Psi_{\sS}$ make
\eqref{Eq:Global_Recovery_Criterion} at $x$ equivalent to
$\zeta>\zeta_{\mathrm{req}}(x)$.
This also covers $\bar a_{\sS}(x)\le1-d_{\sL}$:
the numerator is then zero, and
$\bar a_{\sL}(\zeta\bar\Psi_{\sS}(x))<d_{\sL}$
ensures the criterion for every $\zeta>0$.
Thus recovery holds precisely when
$\zeta>\zeta_{\mathrm{req}}(x)$ for every $x>0$.

For local stability, the small-noise expansion
\eqref{Eq:Reference_Extrinsic_Small_Noise_Proof} gives,
for each fixed $\zeta>0$,
\begin{equation*}
\lim_{x\downarrow0}\frac{\bar\Psi(x)}x
=
\frac{d_{\sL}d_{\sS}}
     {(1-d_{\sL})(1-d_{\sS})}.
\end{equation*}
This limit is less than one when $D<1$ and greater
than one when $D>1$. Hence the zero-error fixed point
is locally attracting in the former case and unstable
in the latter.

\emph{No recovery ($D>1$).}
Here $0<1-d_{\sS}<d_{\sL}$.
As $x\downarrow0$, the numerator of $\zeta_{\mathrm{req}}(x)$
tends to $\bar a_{\sL}^{-1}(1-d_{\sS})>0$, whereas
$\bar\Psi_{\sS}(x)\to0$.
Therefore $\zeta_{\mathrm{req}}(x)\to\infty$ and
$\zeta_{\sC}=\infty$, so recovery fails for every
$\zeta>0$.

\emph{Power-dependent recovery ($D<1<D+\Delta_{\sS}$).}
The extended inverse is continuous on $(0,1]$.
Since $\bar a_{\sS}(x)<1$ and $\bar\Psi_{\sS}(x)>0$,
the function $\zeta_{\mathrm{req}}$ is continuous on $(0,\infty)$.
Moreover,
\begin{equation*}
\bar a_{\sS}(0+)=d_{\sS}<1-d_{\sL},
\qquad
\lim_{x\to\infty}\bar a_{\sS}(x)=0.
\end{equation*}
Thus $\zeta_{\mathrm{req}}(x)=0$ for all sufficiently small
and all sufficiently large $x$.
The condition $D+\Delta_{\sS}>1$ also ensures that
$\zeta_{\mathrm{req}}(x)>0$ for some $x>0$.
Consequently, it attains a finite positive maximum:
\begin{equation*}
0<\zeta_{\sC}
=\max_{x>0}\zeta_{\mathrm{req}}(x)
=\zeta_{\mathrm{req}}(x_{\sC})<\infty
\end{equation*}
for some $x_{\sC}>0$.

For $\zeta>\zeta_{\sC}$, the recovery criterion holds
for every $x>0$.
At $\zeta=\zeta_{\sC}$, equality holds in
\eqref{Eq:Global_Recovery_Criterion} at $x_{\sC}$,
so $\bar\Psi(x_{\sC})=x_{\sC}>0$.
For $\zeta<\zeta_{\sC}$, the criterion fails at the
same $x_{\sC}$.
The convergence result in the preceding subsection
therefore shows that the SE recursion converges to
a nonzero fixed point whenever $0<\zeta\le\zeta_{\sC}$.
Recovery holds exactly when $\zeta>\zeta_{\sC}$.

\emph{All-power recovery ($D+\Delta_{\sS}\le1$).}
For every $x>0$,
\begin{equation*}
\bar a_{\sS}(x)
\le d_{\sS}+\Delta_{\sS}
\le1-d_{\sL}.
\end{equation*}
The extended inverse therefore gives
$\zeta_{\mathrm{req}}(x)=0$ for every $x>0$, and hence
$\zeta_{\sC}=0$.
The recovery criterion holds for every $\zeta>0$,
including at $D+\Delta_{\sS}=1$.
In particular, this includes the allowed boundary case
$D=1$ and $\Delta_{\sS}=0$.

The assumption that $D\ne1$ whenever $\Delta_{\sS}>0$
makes these three cases exhaustive.
\end{proof}

\subsection{Recovery Boundaries for Gaussian-Product Signals}
\label{Appendix:Power_Ratio}

We specialize Theorem~\ref{Theorem:Power_Regimes} to the case where $\bL$
follows the Gaussian-product distribution (see Appendix~\ref{Appendix:Prior_Examples}) and $\bS$ follows the Bernoulli--Gaussian distribution.
Let $r$ be the rank of $\bL$ with $r/m\to\gamma\in(0,1)$, and let $\rho\in(0,1)$
denote the activity probability of the entry prior.
By Appendix~\ref{Appendix:Dimensions},
\begin{equation}
d_{\sL}=\gamma(1+\beta-\beta\gamma),
\qquad d_{\sS}=\rho.
\end{equation}
With
$\Delta_{\sS}(\rho)
=\sup_{x>0}\bar a_{\sS}(x;\rho)-\rho$,
the conditions $D=1$ and $D+\Delta_{\sS}=1$
give the following boundaries:
\begin{align}
\gamma_{\mathrm{loc}}(\rho;\beta)
&=
\frac{1+\beta-\sqrt{(1-\beta)^2+4\beta\rho}}
     {2\beta},
\\
\gamma_{\mathrm{all}}(\rho;\beta)
&=\frac{1+\beta}{2\beta}\nonumber\\
&\quad-
\frac{\sqrt{(1-\beta)^2+
4\beta[\rho+\Delta_{\sS}(\rho)]}}
     {2\beta}.
\end{align}
The zero-error fixed point is locally attracting when
$\gamma<\gamma_{\mathrm{loc}}$.
Recovery holds for every fixed
$\zeta=p_{\sS}/p_{\sL}>0$ when
$\gamma\le\gamma_{\mathrm{all}}$.
For $\gamma_{\mathrm{all}}<\gamma<\gamma_{\mathrm{loc}}$,
recovery instead requires $\zeta>\zeta_{\sC}$,
where $\zeta_{\sC}$ is given in
Theorem~\ref{Theorem:Power_Regimes}.

When $p_{\sS}=p_{\sL}$ (i.e., $\zeta=1$),
the recovery criterion
\eqref{Eq:Global_Recovery_Criterion} becomes
\begin{equation}
\bar a_{\sS}(x;\rho)
+\bar a_{\sL}\bigl(
\bar\Psi_{\sS}(x;\rho);\beta,\gamma
\bigr)<1
\end{equation}
for all $x>0$.

\subsection{Numerical Comparison with State Evolution}
\label{Appendix:Power_Experiments}
We compare the empirical performance of the OAMP algorithm with the SE predictions for the four
configurations labeled A--D in
Fig.~\ref{Fig:Recovery_Phase_Diagram_1}. We fix $p_{\sS}=1$ and vary
$\zeta=p_{\sS}/p_{\sL}$.
The algorithm starts from $\bS^0=\bm0$ and
$\tau_{\sS}^0=p_{\sS}$, with the spectral update
performed first.
For this experiment, we use the input variances and correction
parameters prescribed by SE to avoid finite-sample fluctuations in
their estimates across iterations.
Once an input variance falls below
$10^{-8}\min(p_{\sL},p_{\sS})$ or SE numerically converges to
a positive fixed point, we freeze these parameters and
continue the matrix iterations.

Figure~\ref{Fig:Recovery_Phase_Diagram_2}
shows the empirical error (denoted as $e^t$), defined as the
maximum of the two extrinsic MSEs normalized by
their respective signal powers.
For each $\zeta$, the OAMP curve shows the smallest $e^t$
observed within the 300-iteration budget. The red diamonds show the
SE predictions at iteration 300. Values below $10^{-3}$ are plotted
at $10^{-3}$.
For settings B and C, the vertical lines mark the critical
power ratio $\zeta_{\sC}$ predicted by SE.

\let\originalthebibliography\thebibliography
\renewcommand{\thebibliography}[1]{\originalthebibliography{#1}\interlinepenalty=10000\relax}
\bibliographystyle{IEEEtran}
\bibliography{references}

\end{document}